\documentclass[aps,prd,notitlepage,nofootinbib,superscriptaddress,eqsecnum,floatfix]{revtex4-1}

\usepackage[utf8]{inputenc}
\usepackage{amsmath}
\usepackage{amssymb}
\usepackage{amsfonts}
\usepackage{newtxtext,newtxmath}
\usepackage{bm}
\usepackage{graphicx}

\usepackage[usenames,dvipsnames]{xcolor}
\usepackage{color}
\usepackage[colorlinks=true,linkcolor=blue,urlcolor=blue,citecolor=blue]{hyperref}
\usepackage{booktabs}
\usepackage{slashed}
\usepackage[english]{babel}
\usepackage{dcolumn}
\usepackage{blindtext}
\usepackage{epsfig}
\usepackage{pifont}
\usepackage{dsfont,mathrsfs}
\usepackage{cancel}
\usepackage{bigints}
\usepackage{accents}
\usepackage{soul}
\usepackage{multirow}
\usepackage{subfigure} 
\usepackage{simpler-wick}
\usepackage{float}

\begin{document}

\title{Violation of the Conformal Limit at Finite Density: Insights from Effective Models and Lattice QCD}

\author{Francisco X. Azeredo}\email{francisco.azeredo@acad.ufsm.br}
\affiliation{Departamento de F\'{\i}sica, Universidade Federal de Santa
  Maria,  97105-900 Santa Maria, RS, Brazil}

\author{Arthur E. B. Pasqualotto}\email{ arthur.pasqualotto@acad.ufsm.br}
\affiliation{Departamento de F\'{\i}sica, Universidade Federal de Santa
  Maria,  97105-900 Santa Maria, RS, Brazil}

\author{Bruno S. Lopes}\email{bruno.lopes@acad.ufsm.br}
\affiliation{Departamento de F\'{\i}sica, Universidade Federal de Santa
  Maria,  97105-900 Santa Maria, RS, Brazil}  

\author{ Dyana C. Duarte}\email{ dyana.duarte@ufsm.br}
\affiliation{Departamento de F\'{\i}sica, Universidade Federal de Santa
  Maria, 97105-900 Santa Maria, RS, Brazil}
  
\author{ Ricardo L. S. Farias}\email{ ricardo.farias@ufsm.br}
\affiliation{Departamento de F\'{\i}sica, Universidade Federal de Santa
  Maria, 97105-900 Santa Maria, RS, Brazil}
\affiliation{Center for Nuclear Research, Department of Physics, Kent State University, Kent, OH 44242 USA}   
  
\begin{abstract}
In this work, we discuss recent results obtained with the application of the medium separation scheme (MSS) in different contexts where a clear violation of the conformal limit for the speed of sound at finite density has been observed in Quantum Chromodynamics (QCD). We analyze several scenarios, including QCD at finite isospin density, two-color QCD, and two-flavor color superconductivity. Whenever possible, we compare our findings with lattice QCD (LQCD) results, showing that the Nambu--Jona-Lasinio (NJL) model combined with the MSS provides a consistent description across different regimes of the QCD phase diagram. Our analysis highlights how effective models, when properly regularized, can capture essential nonperturbative features of dense QCD matter, offering complementary insights to lattice simulations.
\end{abstract}

\maketitle

\section{Introduction}

Understanding the behavior of strongly interacting matter at high densities remains one of the central challenges in modern nuclear and particle physics.
In the asymptotically high density regime, perturbative Quantum Chromodynamics (pQCD) for massless quarks becomes approximately invariant under conformal transformations \cite{PARISI1972643}, implying a vanishing $\beta$ function as $\alpha_s(Q) \rightarrow 0$ \cite{Braun:2003rp}, corresponding to a gas of free quarks. In such conformal systems, the trace of the stress--energy tensor vanishes~\cite{Belavin:1984vu,FRIEDAN198693}, yielding $\varepsilon - 3P = 0$ 
and consequently, the conformal limit for the squared speed of sound, $c_s^2 = 1/3$. Whether this value represents a fundamental upper bound at any value of density, or merely the high-density limit of QCD, remains an open question. This issue is particularly relevant for neutron-star physics, where the speed of sound directly constrains the possible combinations of masses and radii~\cite{glendenning2012compact}.
At low densities, Chiral Effective Field Theory (ChEFT) provides reliable guidance~\cite{Drischler:2021kxf}, while pQCD governs the asymptotic regime~\cite{Kurkela:2009gj}. The phenomenologically relevant domain—intermediate densities—lies between these limits, where the violation of the conformal bound and the presence of first-order phase transitions can significantly affect the neutron-star structure~\cite{Tan:2021ahl}. Interpolations constrained by both ChEFT and pQCD indicate that exceeding the conformal limit~\cite{Kurkela:2014vha} is strongly correlated with supporting massive neutron stars (>$2M_\odot$), consistent with other approaches~\cite{Bedaque:2014sqa,Altiparmak:2022bke,Tews:2018kmu}. These observations have also been explored within modified gravity frameworks~\cite{Mendes:2024fel}.

A particularly important subset of systems contained in the dense sector of the QCD phase diagram exhibits a finite isospin density, or an imbalance between the number densities of light quark flavors $u$ and $d$. Classes of compact stellar objects such as neutron stars are characterized by this asymmetry, and investigations regarding the equation of state of QCD matter have been pursued extensively in this context, especially since the detection of gravitational waves by LIGO/Virgo~\cite{LIGOScientific:2017vwq,LIGOScientific:2018cki} and observations made by NICER~\cite{Miller:2019cac,Riley:2019yda,Miller:2021qha,Riley:2021pdl}. From the theoretical point of view, it is possible to study the pure isospin sector in the lattice, as it is not hampered by the sign problem like the nonzero baryon density case. The first results from LQCD were presented in Refs.~\cite{Kogut:2002tm,Kogut:2002zg}, confirming the formation of a Bose--Einstein condensate of charged pions as the isospin chemical potential exceeds the particle's mass, in accordance with the previous works of Refs.~\cite{Son:2000xc,Son:2000by}, where chiral perturbation theory was employed at low density, and perturbative QCD techniques were applied at asymptotically high densities. Efforts were concentrated on the description of the phase diagram, and the transition from the isospin symmetric to the pion superfluid phase was found to be of second order. Posterior lattice implementations focused on different properties of the phase diagram can be found in Refs.~\cite{Kogut:2004zg,deForcrand:2007uz,Cea:2012ev,Detmold:2012wc,Brandt:2017oyy}. Additional studies include the condensation of kaons~\cite{Detmold:2008yn}, multi-pion systems~\cite{Detmold:2008fn}, neutron stars and magnetic field effects~\cite{Endrodi:2014lja}, exotic objects like pion stars~\cite{Brandt:2018bwq}, and computations of the equation of state~\cite{Brandt:2022hwy,Abbott:2023coj,Abbott:2024vhj}.

These last works have also observed the formation of a peak structure in the speed of sound ($c_s^2$) of isospin matter, exceeding the conformal limit. As a quantity directly related to important astrophysical properties such as neutron stars' masses and radii, this result received significant attention in the literature and has since been approached through different perspectives, such as effective models and perturbative techniques. The first reports from Ref.~\cite{Brandt:2022hwy} indicated a peak located at an isospin chemical potential~(definitions found in the literature for the isospin chemical potential $\mu_I$ can vary by a factor two; we adopt $\mu_u = \mu_I/2$ and $\mu_d = -\mu_I/2$ throughout the work, such that $\mu_I = \mu_u - \mu_d$) $\mu_I/m_\pi \sim 1.6$, and a maximum height $c_s^2 \sim 0.56$. This was initially not verified, for example, in Ref.~\cite{Ayala:2023cnt}, where the linear sigma model with quarks (LSMq) and the two-flavor NJL model were employed within the region $\mu_I/m_\pi \in [1.0,1.8]$, corresponding to the available lattice data. Chiral perturbation theory approaches in two and three flavors~\cite{Andersen:2023ivj,Andersen:2023ofv} also did not observe the peak, although the comparison was not direct at increasing density since the relevant degrees of freedom changed. Shortly after, results from the NPLQCD Collaboration~\cite{Abbott:2023coj,Abbott:2024vhj} observed again a violation of the conformal limit at intermediate densities and the formation of a peak structure in $c_s^2$, although its location was shifted to slightly higher densities at $\mu_I/m_\pi \sim 2.5$. Since then, many studies have been successful at describing the aforementioned properties. These include the quark-meson model (or LSMq) in Refs.~\cite{Chiba:2023ftg,Chiba:2024cny,Kojo:2024sca,Ayala:2024sqm,Brandt:2025tkg,Andersen:2025ezj}, the Nambu--Jona-Lasinio model with non-local interactions~\cite{Carlomagno:2024xmi}, medium-dependent couplings in the LSMq and NJL models~\cite{Ayala:2023mms}, holographic QCD~\cite{Kovensky:2024oqx}, perturbative QCD~\cite{Fukushima:2024gmp}, and a quarkyonic picture~\cite{Ivanytskyi:2025cnn}. For a recent review of general features of dense QCD in the lattice, see Ref.~\cite{Itou:2025vcy}. Later in this work, we focus on the description of isospin matter through a two-flavor NJL model with contributions from the medium separation scheme~\cite{Lopes:2025rvn}. This regularization prescription naturally reproduces the peak without the need for additional effects in the model and has shown reliable results also in different but related phenomenologies, as is discussed more extensively throughout the text.

The violation of the conformal limit in dense QCD is not simply a feature of the isospin sector~\cite{Bedaque:2014sqa,Iida:2024irv,Gholami:2024ety,Fukushima:2025ujk}. In Ref.~\cite{Fukushima:2024gmp}, a unified description of the high-density regime based on the Cornwall--Jackiw--Tomboulis formalism was developed, bringing results for the speed of sound and trace anomaly of a two-color diquark superfluid, isospin matter (pion superfluid), as well as the two-flavor color superconducting (2SC) phase. Rather than using the known analytical solutions for the gap functions~\cite{Brown:1999aq,Wang:2001aq,Kanazawa:2009ks,Fujimoto:2023mvc}, the authors considered their full momentum dependence and verified an enhancement as the chemical potentials decreased from the asymptotic limit. For the diquark and pion superfluids, values of $c_s^2$ exceeding the conformal limit were identified along with negative values for the normalized trace anomaly. Corrections due to the gap for the speed of sound of 2SC matter were found to be smaller, and no clear violation of the conformal limit could be observed within the valid perturbative domain. The normalized trace anomaly, nonetheless, still approached negative values.

QCD-like theories provide essential benchmarks for universal properties of strongly interacting matter.
One of the most valuable examples is the two-color QCD (QC$_2$D), where the gauge group SU(2)$_{c}$ is pseudoreal, eliminating the sign problem and enabling lattice simulations across a wide range of densities~\cite{Kogut:2000ek,Kogut:2001na,hands2006deconfinement,ilgenfritz2012two,langfeld2013two,braguta2015two,braguta2014temperature,Braguta:2016cpw,Iida:2024irv}. The phase diagram displays three characteristic regions: hadronic matter, a bosonic diquark superfluid, and a quark–gluon plasma. Theoretical studies using effective models~\cite{kawaguchi2023fate,Adhikari:2018kzh,DSuenaga_2025,Ratti_QC2D} complement lattice insights, and recent progress includes Polyakov-loop effects~\cite{Kurebayashi:2025fcw}; see Ref.~\cite{DSuenaga_2025} for a comprehensive review.

The NJL model has consistently captured the qualitative features of diquark Bose--Einstein condensation (BEC) in QC$_2$D~\cite{Ratti_QC2D}, even though a regularization prescription is required, and thermodynamic consistency becomes crucial. Approaches such as the Renormalization Group~\cite{Gholami:2024diy} and Medium Separation Scheme (MSS)~\cite{Duarte:2018kfd,Farias:2005cr} have proven especially successful. Recent applications of the MSS to QC$_2$D at finite density~\cite{Itou:2023pcl} correctly reproduce the BEC–BCS crossover and the expected peak in $c_s^2$, highlighting the importance of proper disentanglement between vacuum divergences and finite medium contributions.

A particularly appealing aspect of QC$_2$D is that the absence of the sign problem allows for a safe comparison of models with lattice simulations of dense matter, offering valuable benchmarks for model validation. This feature makes QC$_2$D an ideal setting to test how different regularization schemes behave at high density, and to identify which artifacts arise from improper treatments of medium-sensitive quantities. This perspective naturally leads to the broader context of color superconductivity in the realistic three-color QCD, where the correct handling of ultraviolet divergences becomes essential to capture the interplay between pairing dynamics and the thermodynamics of dense quark matter.

Color superconductivity is among the most robust predictions for deconfined quark matter~\cite{Bailin:1983bm,Son:1998uk,Alford:1997zt,Alford:1998mk,Rapp:1997zu,Schmitt:2003xq,Duarte:2018kfd}. At asymptotically high densities the ground state is the Color-Flavor-Locked (CFL) phase, but at neutron stars' densities, the strange quark mass may disfavor CFL, giving rise to the two-flavor color-superconducting phase. This phase is characterized by a diquark condensate that partially breaks the color symmetry from $SU(3)_c$ to $SU(2)_c$, while preserving chiral symmetry, and generates energy gaps typically of the order of tens to hundreds of MeV~\cite{Alford:1997zt,Rapp:1997zu,Rajagopal:2000wf,Rischke:2003mt,Alford:2001dt,Shovkovy:1999mr,Alford:2025jtm}.
The possible existence of deconfined quark matter in the core of neutron stars~\cite{Annala:2019puf,Collins:1974ky} has shed new light on the understanding of strongly interacting matter described by QCD at high densities. Realistic stellar matter must satisfy charge and color neutrality, as well as $\beta$-equilibrium, changing the pairing structure substantially, and in some cases, favoring mixed or spatially inhomogeneous arrangements when surface and Coulomb energies are sufficiently small~\cite{Neumann:2002jm}, and leading to chemical-potential imbalances that introduce Fermi-surface mismatches~\cite{Huang:2002zd,Duarte:2018kfd}. These effects can destabilize pairing, induce gapless modes, or modify the onset of superconductivity~\cite{Shovkovy:1999mr,Huang:2002zd,Alford:2025jtm}.

The approach to the conformal limit, $c_s^2 = 1/3$, plays a central role in understanding the high-density behavior of color-superconducting quark matter. Within the density-functional framework~\cite{Ivanytskyi:2022bjc, Ivanytskyi:2022oxv}, medium-dependent vector and diquark couplings are introduced so that the equation of state (EoS) naturally scales as $P \propto \mu_B^4$ at asymptotically large chemical potentials, thereby recovering conformal behavior. This construction is motivated by earlier discussions on the stiffness of dense matter and the constraints from multi-messenger observations summarized in the review~\cite{Baym:2017whm}, as well as by the insight that perturbative QCD can already constrain the EoS at neutron-star densities through stability and causality arguments~\cite{Komoltsev:2021jzg}. These results also connect to model-agnostic analyses~\cite{Annala:2017llu, Ecker:2022xxj, Altiparmak:2022bke}, which indicate that the speed of sound may peak above the conformal value inside the 2SC phase before decreasing toward $1/3$ at higher densities. Taken together, these works show that the conformal limit provides an essential asymptotic anchor for quark matter, even when strong pairing and nonperturbative interactions dominate at intermediate densities.

In this work, we investigate the violation of the conformal limit at finite density across three distinct scenarios: (i) matter with isospin asymmetry, (ii) two-color QCD with a superconducting pairing channel, and (iii) the more realistic case of three-color QCD including diquark pairing. Our analysis is based on effective models that allow us to explore the thermodynamic behavior of dense matter while providing, when it is possible, a direct comparison with available lattice QCD simulations. A central aspect of this study is the careful treatment of ultraviolet divergences at high densities, for which appropriate regularization schemes are essential to preserve the correct physical behavior of the relevant quantities. By systematically examining these different settings under consistent regularization strategies, we aim to clarify how the interplay between dense-medium effects and model-dependent assumptions influences the violations of the conformal limit.

\section{Medium Separation Scheme}
\label{sec:mss}

The NJL model is non-renormalizable, and any calculation within this framework requires a suitable regularization prescription to render physical quantities finite. In effective model descriptions, medium-dependent contributions frequently appear inside divergent integrals, and a proper separation of such contributions prior to regularization is not a trivial task. A typical example is the momentum integral of the quasiparticle dispersion relation involving the effective quark mass $M$, a chemical potential $\mu_j$, and a generic bosonic condensate $\Delta_i$:
\begin{align}
\int\frac{d^3p}{(2\pi)^3}\sqrt{\left(\sqrt{p^2+M^2}\pm \mu_j\right)^2 + \Delta_i^2}.
\end{align}
{Throughout} 
 this work, the generic chemical potentials $\mu_j$ and condensates $\Delta_i$ appearing in the dispersion relations and thermodynamic potentials acquire different physical interpretations depending on the system under consideration. In particular, $\mu_j$ denotes either the isospin chemical potential $\mu_I/2$ or the quark chemical potential $\mu$ (or, equivalently, the baryon chemical potential $\mu_B = N_c\mu$), while the corresponding condensates are identified as the pion condensate $\Delta_\pi$, the two-color diquark condensate $\Delta_{2c}$, or the 2SC diquark condensate $\Delta_{\text{2SC}}$ in three-color QCD.
The naive procedure commonly found in the literature consists in applying an ultraviolet regulator $\Lambda$ to all momentum integrals. It is well known, however, that such a procedure may introduce spurious medium effects, frequently leading to unphysical thermodynamic behavior. In particular, several studies report that the usual Stefan--Boltzmann limit is not recovered when thermal integrals are truncated by the same cutoff used for vacuum divergences~\cite{Zhuang:1994dw,Costa:2007fy,Costa:2009ae,Xue:2021ldz,Avancini:2020xqe,Pasqualotto:2023hho,Coppola:2024uvz}. Here, the prescription where divergent integrals are regularized without separation of medium effects is referred to as the Traditional Regularization Scheme (TRS).

In contrast, the medium separation scheme proposed in~\cite{Farias:2005cr} and widely discussed in different contexts~\cite{Farias:2016let,Duarte:2018kfd,Das:2019crc,Lopes:2021tro,Azeredo:2024sqc,Pasqualotto:2025kpo,Lopes:2025rvn} implements the physically motivated principle that medium-dependent contributions must be integrated over the full momentum space without restrictions. Vacuum contributions, which are the genuine origin of the ultraviolet divergences, must instead be regularized. By disentangling the thermal, magnetic, and/or density-dependent contributions from the divergent integrals, MSS ensures the correct asymptotic behavior of thermodynamic quantities and avoids the artifacts introduced by a uniform cutoff.

To make the MSS implementation explicit, we now illustrate how the medium and vacuum contributions can be systematically separated in the relevant momentum integrals. For the present analysis, the key integrals are 
\begin{align}
I_{\Delta} & =
\sum_{s=\pm1}\int\frac{d^{3}p}{\left(2\pi\right)^{3}}\frac{1}{E_{\Delta}^{s}},\label{Id3d}\\
I_{M} & = \sum_{s=\pm1}\int\frac{d^{3}p}{(2\pi)^{3}}\frac{1}{E}
\frac{E+s\mu_j}{E_{\Delta}^{s}}, \label{Im3d} 
\end{align}
which appear in the minimization of the thermodynamic potential with respect to $\Delta_i$ and $M$, respectively. In these integrals, we use the definitions $E^s_\Delta = \sqrt{(E+s\mu_j)^2 + \Delta_i^2}$ and $E = \sqrt{p^2+M^2}$.
We can start by rewriting the integral in Equation~\eqref{Id3d} as
\begin{equation}
\int\frac{d^{3}p}{\left(2\pi\right)^{3}}\frac{1}{\sqrt{(E\pm \mu_j)^2 + \Delta_i^2}} = \int\frac{d^{3}p}{\left(2\pi\right)^{3}}\left[\frac{1}{\pi}\int_{-\infty}^{+\infty}\frac{dx}{x^2 + (E\pm \mu_j)^2 + \Delta_i^2}\right].
\label{int1}
\end{equation}
{After} two iterations of the identity

\begin{eqnarray}
 &&\frac{1}{x^{2}+\left(E+s\mu_j\right)^{2}+\Delta_i^{2}}
 =  \frac{1}{x^{2}+p^{2}+M_{0}^{2}} -
\frac{\mu_j^{2}+2sE\mu_j+\Delta_i^{2}+M^{2}-M_{0}^{2}}{\left(x^{2}+E^{2}+
  M_{0}^{2}\right)
  \left[x^{2} +\left(E+s\mu_j\right)^{2}+\Delta_i^{2}\right]},
 \label{ident}
\end{eqnarray}

in the integrand of \eqref{int1}, we obtain
\begin{eqnarray}
\frac{1}{x^{2}+\left(E+s\mu_j\right)^{2}+\Delta_i^{2}} & = & \frac{1}{x^{2}+p^{2}+M_{0}^{2}}
+\frac{A-2sE\mu_j}{\left(x^{2}+p^{2}+M_{0}^{2}\right)^{2}}
+
\frac{\left(A-2sE\mu_j\right)^{2}}{\left(x^{2}+p^{2}+M_{0}^{2}\right)^{3}}
\nonumber \\ & & +
\frac{\left(A-2sE\mu_j\right)^{3}}{\left(x^{2}+
p^{2}+M_{0}^{2}\right)^{3}\left[x^{2}
+\left(E+s\mu_j\right)^{2}+\Delta_i^{2}\right]}.
 \label{r3it}
\end{eqnarray}
{In the} previous expressions, $M_0$ is the effective quark mass evaluated in vacuum, with no medium effects included, and it sets the scale for the MSS approach. We have also defined $A=M_{0}^{2}-M^{2}-\mu_j^{2}-\Delta_i^{2}$. Using the expression for $I_{\Delta}$ in Equation~\eqref{Id3d} together with the integrand obtained in Equation~\eqref{r3it}, the sum over $s$ and the integration over $x$ can be carried out analytically, yielding
\begin{align}
 \int\frac{d^{3}p}{\left(2\pi\right)^{3}}\sum_{s=\pm1} \int_{-\infty}^{+\infty}\frac{dx}{\pi}
 \frac{1}{x^{2}+(E_{\Delta}^{s})^{2}}
 &=
I_{\text{quad}}(M_0)-\frac{\left(\Delta_i^{2}-M_{0}^{2}-2\mu_j^{2}+M^{2}\right)}{2}I_{\text{log}}(M_0)
\nonumber \\ & +
\left[\frac{3\left(A^{2}+4M^{2}\mu_j^{2}\right)}{8}
  -\frac{3\mu_j^{2}M_{0}^{2}}{2}\right]I_{1}-  I_{2},
 \label{temp4}
\end{align}
with the definitions
\begin{eqnarray}
I_{\text{quad}}(M_0) & = &
\int\frac{d^{3}p}{\left(2\pi\right)^{3}}\frac{1}{\sqrt{p^{2}+M_{0}^{2}}},
\label{quad}
\\ I_{\text{log}}(M_0) & = & \int\frac{d^{3}p}{\left(2\pi\right)^{3}}
\frac{1}{\left(p^{2}+M_{0}^{2}\right)^{\frac{3}{2}}},
\label{log}
\\ I_{1}& = & \int\frac{d^{3}p}{\left(2\pi\right)^{3}}
\frac{1}{\left(p^{2}+M_{0}^{2}\right)^{\frac{5}{2}}},
\label{Ifin1}
\\ I_{2}& = &
\frac{15}{32}\sum_{s=\pm1}\int\frac{d^{3}p}{\left(2\pi\right)^{3}}
\int_{0}^{1} dt (1-t)^{2} 
\frac{(A-2sE\mu_j)^{3}}
     {\left[(2sE\mu_j-A)t+p^{2}+M_{0}^{2}\right]^{\frac{7}{2}}}.
 \label{Ifin2}
\end{eqnarray}
{With} this result, one may write Equation~\eqref{Id3d} as
\begin{eqnarray}
I_{\Delta}^{\rm MSS} & = & 2I_{\text{quad}}(M_0)
-\left(\Delta^{2}-M_{0}^{2}-2\mu_j^{2}+M^{2}\right)
I{}_{\text{log}}(M_0)\nonumber \\ &  & +
\left[\frac{3(A^{2}+4M^{2}\mu_j^{2})}{4}-3M_{0}^{2}\mu_j^{2}\right]
I_{1}+2I_{2}.
 \label{Idfinal}
\end{eqnarray}

Similarly, observing that 
\begin{equation}
I_{M} = \sum_{s=\pm1}\int\frac{d^{3}p}{(2\pi)^{3}}\frac{1}{E}
\frac{E+s\mu_j}{E_{\Delta}^{s}} = I_{\Delta} + \sum_{s=\pm1}\int\frac{d^{3}p}{(2\pi)^{3}}
 \frac{1}{E}
\frac{s\mu_j}{E_{\Delta}^{s}}
\end{equation}
and following similar steps for the last integral above, it is possible to write $I_M$ as
\begin{eqnarray}
I_M^{\rm MSS} = I_{\Delta}^{\rm MSS} - 2\mu_j^2 I_{\text{log}}(M_0) - 3A\mu_j^2 I_1 + I_3
\label{Imfinal}
\end{eqnarray}
with the additional definition
\begin{equation}
I_3 = \frac{15}{16}\sum_{s=\pm1}\int\frac{d^{3}p}{(2\pi)^{3}}\int_0^{\infty}dt\frac{t^2}{\sqrt{1+t}}\frac{s\mu_j(A-2s\mu_j E)^3}{E\left[(p^2 + M_0^2)t + (E+s\mu_j)^2 + \Delta_i^2\right]^{\frac{7}{2}}}.
\end{equation}
{Note} that, although the integrals $I_1$, $I_2$, and $I_3$ are written without functional dependence, they depend explicitly on medium parameters such as the effective mass $M$, the diquark condensate $\Delta$, and the chemical potential. Within the MSS, these medium-dependent contributions are finite and can be integrated over the entire momentum range, without the need for an ultraviolet cutoff.
 The expressions for the gap equations involving these integrals were obtained explicitly in Ref.~\cite{Duarte:2018kfd}.
The thermodynamic potential, on the other hand, has integrals of the form {(the subtracted term $2E_0$, which ensures that the pressure vanishes in the vacuum, is crucial for the cancellation of the quartic divergence within the MSS framework)}
\begin{align}\label{EqVeffTRS}
I_{\Omega} = \int \frac{d^3 p}{(2\pi)^3}\left(E_{\Delta}^{+} + E_{\Delta}^{-} - 2E_0\right) \, ,
\end{align}
which is given in the MSS approach as 
\begin{align}\label{EqVeffMSS}
I_{\Omega}^{\rm MSS} &=\tilde{\mathcal{M}} I_{\rm quad}(M_0) - \frac{1}{4}\left(\tilde{\mathcal{M}}^2 - 4\mu_j^2 \Delta_i^2\right)I_{\rm log}(M_0) \nonumber\\
&+ \int \frac{d^3 p}{(2\pi)^3} \left( \frac{\tilde{\mathcal{M}}^2 - 4\mu_j^2 \Delta_i^2}{4E_{0}^3} - \frac{\tilde{\mathcal{M}}}{E_{0}} - 2E_{0} + E_{\Delta}^+ + E_{\Delta}^-  \right) \, ,
\end{align}
in which $\tilde{\mathcal{M}} = \Delta_i^2 + M^2 - M_0^2$ and $E_{0} = \sqrt{p^2 + M_0^2}$.

{It is worth emphasizing that the medium separation scheme is not based on an arbitrary rearrangement of integrals, but on a well-defined algebraic manipulation of the integrals on the gap equations, performed in terms of a vacuum mass scale $M_0$. This scale corresponds to the constituent quark mass in vacuum and is uniquely fixed through the parametrization of the model by fitting vacuum observables. Its use is physically motivated: in effective field theories, ultraviolet divergences originate from vacuum contributions, whereas medium effects do not introduce additional divergences.
In principle, one could perform the separation using an arbitrary scale; however, without a physically motivated prescription to fix such a scale independently of medium effects, the procedure would become ambiguous. Thus, once the regularization scheme and the parametrization are fixed, the MSS provides a well-defined and physically supported separation between divergent vacuum terms and finite medium contributions. Moreover, alternative algebraic rearrangements that preserve the requirement that only vacuum divergences are regularized lead to equivalent finite medium-dependent results, ensuring the robustness of the MSS.}

\section{Vacuum Regularization}

In Equations~\eqref{Idfinal} and~\eqref{Imfinal}, the medium contributions have been completely removed from the divergent integrals, and only $I_{\text{quad}}$ and $I_{\text{log}}$ still require a regularization. They are expressed, however, in terms of the effective quark mass $M_0$, whose value is fixed once the model parameters are chosen in the vacuum, given its linear relation with the quark condensate, $M = m_c - 2G_S\langle\bar{q}q\rangle$ {(more details about the model parametrizations are shown in the next sections)}.
In this section, we discuss two approaches to vacuum regularization: the use of a 3D momentum cutoff (3D) and the Pauli--Villars (PV) method.
Once again, the TRS refers to the results obtained via regularization without the medium separation procedure described in the last section, for both these methods.

The 3D cutoff is the most commonly employed scheme in the literature, which consists simply in modifying the upper integration limit from infinity to a finite value $\Lambda_{\rm 3D}$:
\begin{equation}
\int\frac{d^{3}p}{(2\pi)^{3}} \to \int_0^{\Lambda_{\rm 3D}}\frac{d
p~p^2}{2\pi^{2}}
\end{equation}
{This} sharp cutoff of high momenta introduces $\Lambda_{\rm 3D}$ as a new parameter of the model, together with the current quark mass $m_c$ and the scalar coupling $G_S$.

Pauli--Villars regularization, on the other hand, consists in replacing the integrand of the divergent integral with a linear combination of fictitious mass terms carefully chosen to ensure the cancellation of ultraviolet divergences. That is,
\begin{align}
\int_{0}^{\infty} dp \, f(p,M) \to \int_{0}^{\infty} dp \sum_{i=1}^{3} c_i \, f(p,M_i) \, ,
\end{align}
with modified masses defined as 
\begin{equation}
M_i^2 = M^2+ \alpha_i \Lambda_{\rm PV}^2,
\end{equation}
where $\Lambda_{\rm PV}$ denotes the regulator associated with Pauli--Villars scheme. The coefficients $c_i$ and $\alpha_i$ are chosen so that the divergent terms cancel in the asymptotic limit. In this work, we use 

\begin{align}
    c_1 &= 1, & \alpha_1 = 0,\nonumber \\ c_2 &= -2,& \alpha_2 = 1,\nonumber \\ c_3 &= 1, & \alpha_3 = 2.
\end{align}
{With} these values, the PV-regularized forms of the divergent integrals become 
\begin{align}
I_{\rm quad,PV}(M_0) =& \frac{1}{4\pi^2}\left[M_0^2 \log(M_0)-(M_0^2+\Lambda_{\rm PV}^2)\log(M_0^2+\Lambda_{\rm PV}^2)\right.\nonumber\\
&\left.+\frac{(M_0^2+2\Lambda_{\rm PV}^2)}{2}\log(M_0^2+2\Lambda_{\rm PV}^2)\right],\nonumber\\
I_{\rm log,PV}(M_0) &= \frac{1}{2\pi^2}\left[-\log(M_0) + \log(\Lambda_{\rm PV}^2 + M_0^2) - \frac{\log(2\Lambda_{\rm PV}^2 +M_0^2)}{2}\right].  
\end{align}

In the next sections, we examine how the choice of the regularization scheme influences the violation of the conformal limit across three distinct scenarios: isospin-asymmetric matter, two-color QCD with pairing, and three-color QCD including diquark condensation.
\section{QCD at Finite Isospin Density}
\label{sec:isospin}

Discussions regarding the formalism of the NJL model with an isospin asymmetry have been realized extensively throughout the years, with applications to a variety of physical systems~\cite{Toublan:2003tt,Frank:2003ve,Barducci:2004tt,He:2005sp,He:2005nk,Lawley:2005ru,He:2006tn,Ebert:2005wr,Ebert:2005cs,Shao:2006gz,Mukherjee:2006hq,Andersen:2007qv,Sun:2007fc,Abuki:2008wm,Mu:2010zz,Xia:2013caa,Ebert:2016hkd,Khunjua:2017khh,Khunjua:2018sro,Khunjua:2018jmn,Duarte:2018kfd,Khunjua:2019lbv,Khunjua:2019ini,Avancini:2019ego,Lu:2019diy,Khunjua:2020hbd,Khunjua:2020xws,Lopes:2021tro,Khunjua:2021oxf,Liu:2021gsi,Ayala:2023mms,Liu:2023uxm,Khunjua:2024kdc,Carlomagno:2024xmi,Lopes:2025rvn,Klimenko:2025kqi}. We recapitulate the essential details here, working within a two-flavor description which is representative of the general case as shown in Ref.~\cite{Lopes:2025rvn} and as expected since the isospin imbalance is originated in the light quark sector. The Lagrangian density is defined by
\begin{align}
\mathcal{L} &= \bar{\psi}\left(i\gamma^\mu \partial_\mu - m_c\right)\psi + G_S \left[ \left(\bar{\psi}\psi\right)^2 + \left(\bar{\psi}i\gamma_5 \vec{\tau} \psi\right)^2 \right] \, ,
\end{align}
for quark spinors $\psi = (u,d)^T$ in flavor space, current quark mass $m_c$, scalar/pseudoscalar interaction coupling $G_S$, and Pauli matrices vector $\vec{\tau} = (\tau_1,\tau_2,\tau_3)$. The partition function at finite temperature $T$, isospin and baryon chemical potentials $\mu_I$ and $\mu_B$, respectively, is given by
\begin{align}
Z (T,\mu_B,\mu_I) = \int \left[d\bar{\psi}\right]\left[d\psi\right] \exp \left[ - \int_0^\beta d\tau \int d^3 x \left( \mathcal{L} + \bar{\psi}\hat{\mu}\gamma_0\psi \right) \right] \, ,
\end{align}
where $\hat{\mu} = \textrm{diag}\left(\mu_u,\mu_d\right)$ is the chemical potential matrix. The chemical potential of quark flavors $u$ and $d$ are related to $\mu_I$ and $\mu_B$ through
\begin{align}
&\mu_u = \frac{\mu_B}{3} + \frac{\mu_I}{2} \, , \nonumber \\
&\mu_d = \frac{\mu_B}{3} - \frac{\mu_I}{2} \, ,
\end{align}
such that $\mu_B / 3 = (\mu_u + \mu_d) / 2$ and $\mu_I = \mu_u - \mu_d$. We consider only the case $T = \mu_B = 0$ from here onward. Anticipating the condensation of charged pions, it is useful to work on the basis ($\tau_+,\tau_-,\tau_3$) defined through the combinations
\begin{align}
\tau_{\pm} = \frac{(\tau_1 \mp i\tau_2)}{\sqrt{2}} \, .
\end{align}
{The} Lagrangian density can then be written as
\begin{align}
\mathcal{L} = \bar{\psi}\left(i\gamma^\mu \partial_\mu - m_c\right)\psi + G_S \left[ \left(\bar{\psi}\psi\right)^2 + \left(\bar{\psi}i\gamma_5 \tau_3 \psi\right)^2 + 2 \left(\bar{\psi}i\gamma_5 \tau_{+}\psi\right)\left(\bar{\psi}i\gamma_5 \tau_{-}\psi\right)\right] \, .
\end{align}
{The} finite isospin chemical potential explicitly breaks the isospin symmetry SU(2) down to the subgroup U(1)$_{I_3}$, with the third component of the isospin charge being the generator~\cite{Mu:2010zz}. It then suffices to consider $\left\langle \bar{\psi}i\gamma_5 \tau_3 \psi \right\rangle = 0$ as there is no condensation of neutral pions, while charged condensates $\left\langle \bar{u}i\gamma_5 d \right\rangle = \left\langle \bar{d}i\gamma_5 u \right\rangle^{*} \neq 0$ may differ by an arbitrary phase and further break the residual symmetry. Working in the mean-field approximation, the scalar condensate $\sigma$ and pion condensate $\Delta_\pi$ are introduced as
\begin{align}
\sigma &= - 2 G_S \left\langle \bar{\psi} \psi \right\rangle \, , \nonumber\\
\sqrt{2}\pi_+ &= - 2\sqrt{2} G_S \left\langle \bar{\psi} i\gamma_5 \tau_+ \psi \right\rangle = \Delta_\pi e^{i \theta} \, , \nonumber\\
\sqrt{2}\pi_- &= - 2\sqrt{2} G_S \left\langle \bar{\psi} i\gamma_5 \tau_- \psi \right\rangle = \Delta_\pi e^{-i \theta} \, .
\end{align}
{The phase} $\theta$ indicates the direction of symmetry breaking and can be taken as zero without loss of generality~\cite{Mu:2010zz}. Finally, the effective Lagrangian can be written as
\begin{align}
\mathcal{L} + \bar{\psi}\hat{\mu}\gamma_0\psi  = \bar{\psi}\left(i\gamma^\mu \partial_\mu - m_c - \sigma - \Delta_\pi i \gamma^5 \tau_1 + \frac{\mu_I}{2}\gamma^0\tau_3\right)\psi - \frac{\sigma^2 + \Delta_\pi^2}{4G_S} \, .
\end{align}
{Then}, in the mean-field approximation, the thermodynamic potential at finite $\mu_I$ is given by
\begin{align}\label{eqn:iso_omega}
\Omega = \frac{\sigma^2 + \Delta_\pi^2}{4 G_S} - 2 N_c \int \frac{d^3 p}{(2\pi)^3}\left(E_{\Delta}^{+} + E_{\Delta}^{-}\right) \, ,
\end{align}
with $N_c = 3$ being the number of colors and
\begin{align}
E_{\Delta}^{\pm} = \sqrt{\left(E \pm \frac{\mu_I}{2} \right)^2 + \Delta_\pi^2}
\end{align}
the dispersion relations, where $E = \sqrt{\vec{p}^2 + M^2}$, and $M = m_c + \sigma$ is the effective quark mass generated by spontaneous chiral symmetry breaking. The physical values of condensates $\sigma$ and $\Delta_\pi$ minimize the effective potential, thus satisfying the gap equations
\begin{align}
\frac{\partial \Omega}{\partial \sigma} = \frac{\partial \Omega}{\partial \Delta_\pi} = 0 \, .
\end{align}
{The} integral involving the energies in Equation~\eqref{eqn:iso_omega} is divergent and must be regularized. As previously motivated, we consider both the traditional regularization scheme and the medium separation scheme. For the latter, the potential is written according to Ref.~\cite{Avancini:2019ego} through the steps described in Section~\ref{sec:mss}.
\par As for the thermodynamic quantities of interest, we define the pressure $P$, isospin density $n_I$, energy density $\varepsilon$, and speed of sound $c_s^2$
\begin{alignat}{2}
&P &&= - [\Omega(\mu_I) - \Omega(\mu_I = 0)] \, , \nonumber \\
&n_I &&= \frac{\partial P}{\partial \mu_I} \, , \nonumber \\
&\varepsilon &&= - P + \mu_I n_I \, , \nonumber \\
&c_s^2 &&= \frac{\partial P}{\partial \varepsilon} \, .
\end{alignat}
{The} parametrization adopted in the isospin dense case is shown in Table~\ref{tab:iso_parameters} for both the 3D-cutoff and Pauli--Villars regularization prescriptions. These sets were shown to yield good results for the speed of sound in Ref.~\cite{Lopes:2025rvn}, especially compared to the available lattice data from Refs.~\cite{Abbott:2023coj,Abbott:2024vhj}. They correspond to the physical values in vacuum of the pion decay constant $f_\pi = 92.4$ MeV, pion mass $m_\pi = 139.57$ MeV, and light quark condensate $\left\langle\bar{u}u\right\rangle^{1/3} = - 250$ MeV. The resulting effective quark mass in vacuum is $M_0 = 288.1$ MeV and $M_0 = 236.9$ MeV for the 3D and PV regularizations, respectively. {Although a different choice for the parameter sets may yield slightly different results, the qualitative picture should remain. In this case, the pion mass is fixed to $m_\pi = 139.57$ MeV since this value is used in Ref.~\cite{Abbott:2024vhj}, to which our results for the speed of sound are compared.}

\begin{table}[H]
	\caption{Parameter sets for the SU(2) NJL model at finite isospin density.}
	\begin{center}
	\setlength{\tabcolsep}{17.6mm}
		\begin{tabular}{ccc}
			\toprule[0.5pt]
			\textbf{Parameter} & \textbf{PV} & \textbf{3D} \\ 
			\midrule[0.25pt]
			$G_S\Lambda^2$        & $2.763$ & $2.029$ \\ 
			$\Lambda$ (MeV)     & $863.3$ & $669.5$ \\ 
			$m_c$ (MeV)         & $5.205$ & $5.226$ \\ 
			\bottomrule[0.5pt]
		\end{tabular}
	\end{center}
	\label{tab:iso_parameters}
\end{table}
\begin{figure}[h]
\includegraphics[width=0.65\linewidth]{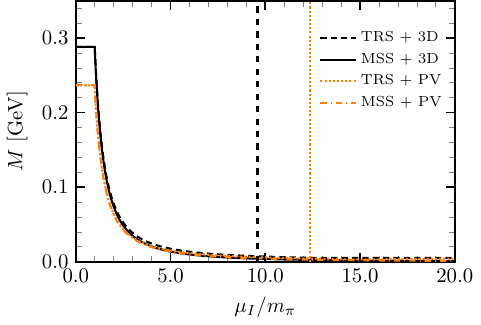}
\caption{Effective quark mass $M$ as a function of the normalized isospin chemical potential $\mu_I/m_\pi$. Dashed, solid, dotted, and dot-dashed lines correspond to TRS + 3D, MSS + 3D, TRS + PV, and MSS + PV, respectively. {The vertical black dashed line indicates the scale $\Lambda_{\text{3D}}$, while the vertical orange dotted line represents the corresponding Pauli--Villars (PV) cutoff, $\Lambda_{\text{PV}}$.}}
\label{fig:iso_mass}
\end{figure}
\par In Figure~\ref{fig:iso_mass}, we show the behavior of the effective quark mass $M$ as a function of the isospin chemical potential $\mu_I$. It assumes the constant vacuum value $M_0$ before the onset of pion condensation at $\mu_I = m_\pi$ and decreases thereafter, partially restoring chiral symmetry. Values of $M$ are slightly higher for the 3D-cutoff method compared to the Pauli--Villars case, which is more evident in the isospin symmetric region at $\mu_I < m_\pi$. More noticeably, the differences between results from the TRS and MSS are small for both 3D and Pauli--Villars regularizations. Although this is true for this quantity, we verify afterwards that an improper choice of the regularization prescription has major consequences for the behavior of thermodynamic quantities. It is also important to mention that results for values of the quark chemical potential $|\mu| = \mu_I / 2$ above the regularization scales $\Lambda_{\rm 3D}$ and $\Lambda_{\rm PV}$ are an extrapolation and thus must be examined with caution. These limits correspond to $\mu_I / m_\pi < 9.59$ in the 3D case and $\mu_I / m_\pi < 12.37$ in the PV case. This point is further elaborated on in the following discussions.
\begin{figure}[b]
\includegraphics[width=0.65\linewidth]{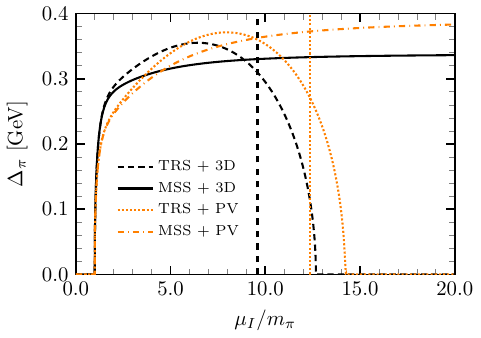}
\caption{Pion condensate $\Delta_\pi$ as a function of the normalized isospin chemical potential $\mu_I/m_\pi$. The line styles follow the same convention as in Figure~\ref{fig:iso_mass}. {The vertical black dashed line indicates the scale $\Lambda_{\text{3D}}$, while the vertical orange dotted line represents the corresponding Pauli--Villars (PV) cutoff,~$\Lambda_{\text{PV}}$.}}
\label{fig:iso_cond}
\end{figure} 
\par In Figure~\ref{fig:iso_cond}, we present the pion condensate $\Delta_\pi$ as a function of the isospin chemical potential. The onset of the condensed phase at $\mu_I = m_\pi$ can be clearly seen, and in all cases considered, the gap is a rapidly increasing function of $\mu_I$ near this point. As the chemical potential increases, the differences start to become clear. For both 3D and PV regularizations, the TRS prescription yields a decreasing gap function at intermediate to high densities, until it reaches zero again, signaling an end to the condensed phase. In the MSS, however, both regularization methods yield an increasing condensate $\Delta_\pi$ throughout the whole region considered in $\mu_I$. This fact turns out to be deeply important for a proper description of the speed of sound, as seen later on. In Ref.~\cite{Fukushima:2024gmp}, for example, within a pQCD description of the superfluid phase, the presence of the gap significantly alters the behavior of $c_s^2$ and is necessary to properly identify a violation of the conformal limit. Ref.~\cite{Abbott:2024vhj} also estimates the condensate at high densities through a comparison of the lattice pressure with pQCD, determining that it still contributes significantly in this sector. Thus, although it may involve an extrapolation of the region of validity of the NJL model as we briefly described, the persistence of the gap at high densities seems to be in accordance with other approaches and is a strong argument in favor of the MSS.
\begin{figure}[b]
\includegraphics[width=0.65\linewidth]{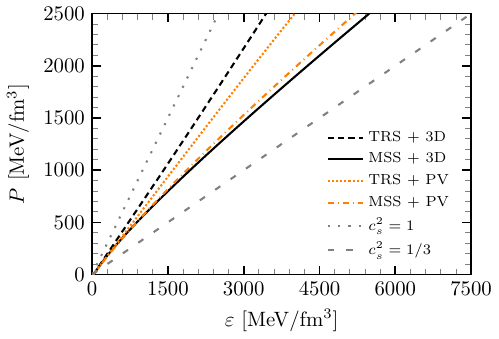}
\caption{Equation of state $P \times \varepsilon$ of isospin dense matter. The line styles follow the same convention as in Figure~\ref{fig:iso_mass}, and guidelines corresponding to constant values of the speed of sound, $c_s^2 = 1/3$ and $c_s^2 = 1$, are also shown.}
\label{fig:iso_eos}
\end{figure} 
\par Figure~\ref{fig:iso_eos} concerns the equation of state of isospin dense matter in the NJL model. We first note that the MSS description yields similar results regardless of the regularization method (3D or PV) employed. Compared to the TRS, it also results in a softer equation of state (smaller pressure for a determined value of the energy density) as $\varepsilon$ increases. Once again, these differences stem from the fact that important medium contributions are being suppressed by the regulators in the traditional scheme. Also shown are the guidelines for constant values of the speed of sound $c_s^2 = 1/3$ and $c_s^2 = 1$. All curves lie in between these lines within the presented density regime, which does not include the problematic region where $\Delta_\pi \to 0$ in the TRS. This sector is discussed in sequence from the point of view of the speed of sound.
\begin{figure}[t]
\includegraphics[width=0.65\linewidth]{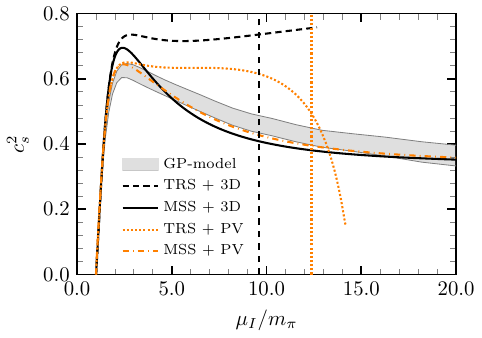}
\caption{Speed of sound squared $c_s^2$ as a function of the normalized isospin chemical potential $\mu_I/m_\pi$. The line styles follow the same convention as in Figure~\ref{fig:iso_mass}. {The vertical black dashed line indicates the scale $\Lambda_{\text{3D}}$, while the vertical orange dotted line represents the corresponding Pauli--Villars (PV) cutoff, $\Lambda_{\text{PV}}$.} }
\label{fig:iso_cs2}
\end{figure} 
\par Finally, Figure~\ref{fig:iso_cs2} shows the behavior of the speed of sound as a function of the isospin chemical potential. In addition to all model predictions previously described, we also present the data from the Gaussian Process (GP) model of Ref.~\cite{Abbott:2024vhj}. It combines results from chiral perturbation theory at low densities, lattice QCD in the intermediate region, and pQCD at high densities, avoiding problems due to finite-lattice-size spacing effects and uncontrolled systematics. The data reveal that at the onset of pion condensation, $c_s^2$ becomes nonzero, rapidly increasing until reaching a maximum and decreasing to the conformal limit $c_s^2 = 1/3$ from above thereafter. It is then this quantity that demonstrates the most remarkable differences between TRS and MSS descriptions. Note that only within the MSS are all these features reproduced, showing a clear peak structure and a convergence to the conformal limit regardless of the regulator applied. The TRS, on the other hand, does not describe a clear peak and even fails at assigning finite values of $c_s^2$ after the point at which $\Delta_\pi$ returns to zero. While it is true that this latter point happens outside the commonly adopted region of validity of the NJL model (i.e., at chemical potentials above the regularization scale), the TRS is still unable to reproduce the data near the peak and shortly after, which is well within the validity domain. In short, the medium contributions taken into account due to the medium separation scheme are thus crucial for a correct description of the overall behavior. Within the MSS, the Pauli--Villars regularization seems to more closely resemble the changes in curvature shown by the reference data compared to the 3D case, although both methods yield good qualitative agreement. Ref.~\cite{Lopes:2025rvn} includes further discussions on this topic, along with comparisons to results from the quark--meson models of Refs.~\cite{Andersen:2025ezj,Brandt:2025tkg}, which are only captured in the NJL model by the MSS.
%

%

\section{Two-Color QCD}
\label{sec:qc2c}

In this section, we summarize the SU(2)$_f$ NJL framework for two-color QCD. The QC$_2$D formulation is constructed in close analogy to the more familiar three-color counterpart but incorporates an additional diquark channel that captures the onset of baryonic superfluidity through diquark condensation. The corresponding Lagrangian density reads~\cite{Ratti_QC2D}:

\begin{align}
\mathcal{L} &= \bar \psi (i\gamma^\mu\partial_\mu-m_c)\psi+
G_S[(\bar\psi\psi)^2 +(\bar\psi i \gamma_5 \vec\tau
  \psi)^2]  \label{NJL_diq}
  \nonumber\\ & \ \ \ \ + G_D(\bar \psi i \gamma_5 \tau_2 t_2 C
\bar\psi^T)(\psi^T C i \gamma_5 \tau_2 t_ 2 \psi),
\end{align}
where $G_S$ and $G_D$ denote the scalar--pseudoscalar and diquark couplings, respectively.
The matrices $\tau_2$ and $t_2$ correspond to the second Pauli matrix in the flavor and color space, $\vec{\tau} = (\tau_1,\tau_2,\tau_3)$ spans the flavor generators, and $C = i \gamma^2 \gamma^0$ is the charge conjugation matrix.
A Fierz transformation in color space relates the interaction channels, yielding $ G_S= G_D\equiv G$~\cite{Ratti_QC2D,Sun:2007fc}. 

Because the NJL model is non-renormalizable, a regularization procedure is required to obtain physical vacuum quantities,  which are subsequently used to fix the model parameters. These are chosen to reproduce the properties of the lightest pseudoscalar mode of the theory (the analogue of the pion in real QCD), namely its mass $m_{\pi}$, decay constant $f_{\pi}$, and the quark condensate in the vacuum, $\langle\bar{q}q\rangle_0$. 

Since two-color QCD is primarily a theoretically guided framework, some freedom remains in selecting these vacuum inputs, and correspondingly in the parameter sets used in phenomenological applications. 
In this work, we adopt the same $N_c$ scaling employed in~\cite{Brauner:2009gu,Duarte:2015ppa,Pasqualotto:2025kpo}, whereby two-color quantities $f_\pi$ and $\langle \bar qq\rangle_0$ are rescaled by the factors $\sqrt{2/3}$ and $2/3$, respectively. The $N_c =2$ inputs are therefore obtained from their $N_c = 3$ counterparts, $f_\pi = 92.4$ MeV and $-\langle \bar qq \rangle^{1/3} = 250$ MeV, while the pion mass remains unchanged because its value is independent of $N_c$. This leads to the vacuum inputs: $m_\pi = 140$ MeV, $f_\pi = 75.45$ MeV, and $\langle \bar qq\rangle^{1/3} = -218$ MeV.
Since each parametrization is determined by fitting these vacuum quantities through the regularized equations, different regularization schemes naturally yield different values for the coupling $G$, current quark mass $m_c$, and momentum of the ultraviolet scale $\Lambda$. 
Consequently, these values also affect the effective quark mass in the vacuum, yielding $M_0 = 301$ MeV and $237$ MeV for the 3D and PV parameter sets, respectively.
The numerical values employed in this section are summarized in Table~\ref{tab:qc2c_parameters}.

\begin{table}[H]
	\caption{Parameter sets for the SU(2) NJL model in QC$_2$D.}
	\begin{center}
	\setlength{\tabcolsep}{18.0mm}
		\begin{tabular}{ccc}
			\toprule[0.5pt]
			\textbf{Parameter} & \textbf{PV} & \textbf{3D} \\ 
			\midrule[0.25pt]
			$G\Lambda^2$        & $4.16$ & $3.107$ \\ 
			$\Lambda$ (MeV)     & $860$  & $657$ \\ 
			$m_c$ (MeV)         & $5.30$ & $5.30$ \\ 
			\bottomrule[0.5pt]
		\end{tabular}
	\end{center}
	\label{tab:qc2c_parameters}
\end{table}

The thermodynamic potential of two-color QCD within the NJL framework is given by~\cite{Sun:2007fc}
\begin{eqnarray}
\Omega &=& \frac{(M-m_c)^2 + \Delta_{2c}^2}{4G} - N_c N_f\sum_{s = \pm 1}\int\frac{d^3p}{(2\pi)^3}E_{\Delta_{2c}}^s\nonumber\\
& & -2N_c N_f T\sum_{s = \pm 1}\int\frac{d^3p}{(2\pi)^3}\ln{\left[1 + \exp\left({-{E_{\Delta_{2c}}^s/T}}\right)\right]}, \label{Omega_2c}
\label{OmegaT}
\end{eqnarray}
where the modified dispersion relation is $E_{\Delta_{2c}}^s = \sqrt{(\sqrt{p^2 + M^2} + s\mu )^2 + \Delta_{2c}^2}\,$. 
In this relation, $\mu$ is the quark chemical potential, $M$ denotes the effective quark mass, and $\Delta_{2c}$ represents the diquark condensate. 

Both quantities are determined self-consistently by minimizing the thermodynamic potential~\eqref{Omega_2c} with respect to $\Delta_{2c}$ and $M$:
\begin{eqnarray}
\frac{\partial\Omega}{\partial\Delta_{2c}}= \frac{\partial\Omega}{\partial M}
=0,
\end{eqnarray}
which define the corresponding gap equations. The MSS version of these equations is obtained by following the steps of Section~\ref{sec:mss} and Ref.~\cite{Duarte:2018kfd}.

By implementing both 3D-cutoff and Pauli--Villars regularization schemes, each considered with and without the proper treatment of medium contributions via the MSS, we are able to compare the behavior of the order parameters $M$ and $\Delta_{2c}$, as functions of the quark chemical potential $\mu$.
In the following, we denote by traditional regularization scheme the prescription in which medium effects are not separated from the divergent vacuum integrals, in contrast to the MSS approach, where the separation is explicitly performed.

\begin{figure}[h]
    \includegraphics[width=0.65\linewidth]{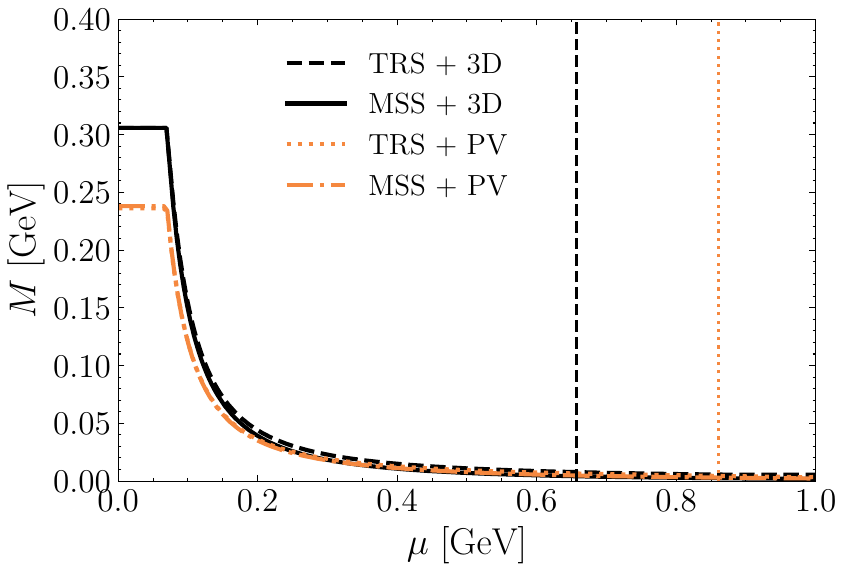}
    \caption{{Effective}
 mass $M$ as a function of quark chemical potential $\mu$. {The vertical black dashed line indicates the scale $\Lambda_{\text{3D}}$, while the vertical orange dotted line represents the corresponding Pauli--Villars (PV) cutoff, $\Lambda_{\text{PV}}$.} The line styles follow the same convention as in Figure~\ref{fig:iso_mass}.}
    \label{fig:nc2_Meff}
\end{figure}

\begin{figure}[t]
    \includegraphics[width=0.65\linewidth]{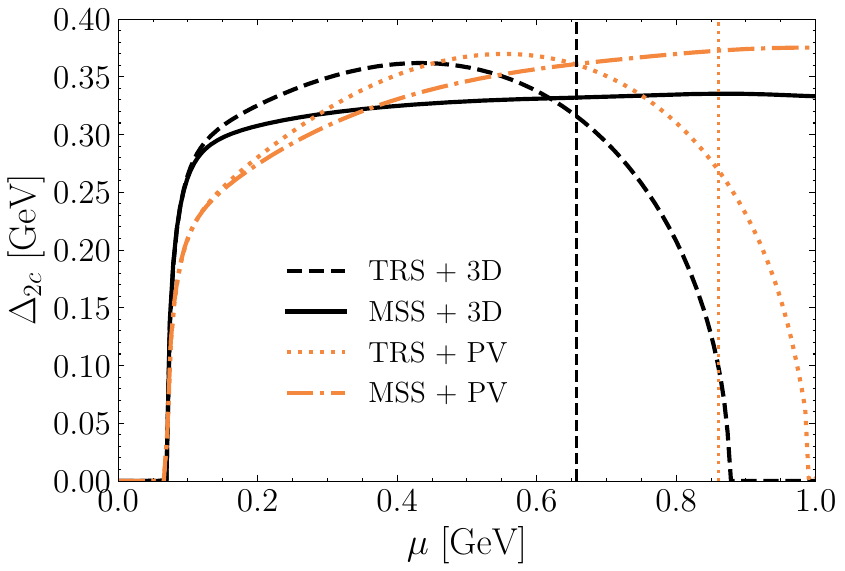}
    \caption{{Diquark} 
 condensate $\Delta_{2c}$ as a function of quark chemical potential $\mu$. {The vertical black dashed line indicates the scale $\Lambda_{\text{3D}}$, while the vertical orange dotted line represents the corresponding Pauli--Villars (PV) cutoff, $\Lambda_{\text{PV}}$.} The line styles follow the same convention as in Figure~\ref{fig:iso_mass}.}
    \label{fig:nc2_Delta}
\end{figure}

The resulting $\mu$-dependence of the effective quark mass $M$ and diquark condensate  $\Delta_{2c}$ for each regularization scheme is displayed in Figures~\ref{fig:nc2_Meff} and~\ref{fig:nc2_Delta}, respectively.

As shown in Figure~\ref{fig:nc2_Meff}, the behavior of the effective mass as a function of $\mu$ is only weakly sensitive to the MSS procedure, both in the 3D-cutoff and Pauli--Villars case. In particular, the TRS and MSS lead to the same qualitative pattern, differing primarily in the quantitative value of the mass plateau in the region  $0 \leq \mu \leq m_\pi/2$, which reflects the specific choice of vacuum regularization. As the chemical potential exceeds $m_{\pi}/2$, the effective mass exhibits a sharp decrease, characteristic of a second-order phase transition,  and subsequently converges toward the current quark mass $m_c$ at large values of $\mu$.
In contrast to the effective mass, the diquark condensate $\Delta_{2c}$ exhibits a pronounced sensitivity to the MSS procedure. Figure~\ref{fig:nc2_Delta} displays the $\mu$-dependence of $\Delta_{2c}$ for all regularization schemes considered.
Within the traditional, non-separated TRS prescription, employing either the 3D-cutoff or the Pauli–Villars regularization, the diquark condensate follows a similar qualitative pattern: a relatively steep initial increase with $\mu$, followed by the development of a maximum and a subsequent rapid decrease to zero. In the TRS framework with a 3D cutoff, $\Delta_{2c}$ reaches its maximum at $\mu\approx 0.4$ GeV and vanishes around $\mu\approx 0.87$ GeV. For the Pauli--Villars TRS case, the maximum is shifted to $\mu\approx 0.6$ GeV, with the condensate disappearing at $\mu\approx 1$ GeV.

A qualitatively different behavior emerges once the medium separation scheme is implemented. Within the MSS, both the 3D-cutoff and Pauli–Villars regularizations lead to an almost identical qualitative evolution of $\Delta_{2c}$: the condensate increases monotonically throughout the explored chemical potential range, with no indication of a decrease for $\mu\approx 1$ GeV. In the Pauli--Villars case, $\Delta_{2c}$ reaches a plateau of approximately $0.37$ GeV, whereas the 3D-cutoff regularization yields a maximum value of $\Delta_{2c}\approx 0.33$ GeV, followed only by a very mild decrease. The monotonic behavior of the diquark condensate obtained within the MSS is compatible with results from chiral perturbation theory at low chemical potentials~\cite{Ratti_QC2D}, and qualitatively consistent with lattice simulations performed with heavy pion masses~\cite{Iida:2024irv}.

Finally, the Pauli–Villars curves obtained within the MSS and TRS closely track each other at low chemical potentials, remaining nearly indistinguishable up to $\mu\approx 0.17$ GeV before departing from one another. In contrast, for the 3D-cutoff regularization, this separation already occurs at smaller values, around $\mu\approx 0.1$ GeV.

\begin{figure}[t]
    \includegraphics[width=0.65\linewidth]{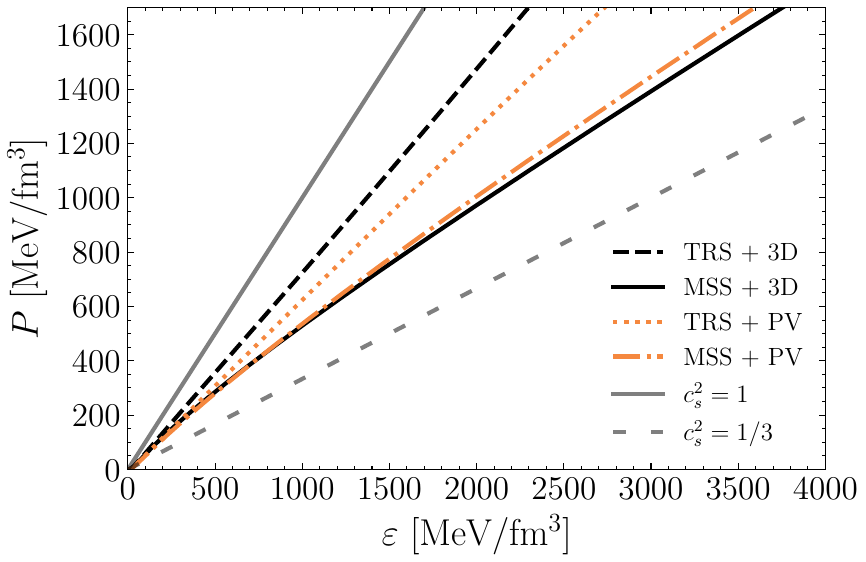}
    \caption{Equation of state (EoS) $P\times \varepsilon$ for two-color QCD. The line styles follow the same convention as in Figure~\ref{fig:iso_mass}, and guidelines corresponding to constant values of the speed of sound, $c_s^2 = 1/3$ and $c_s^2 = 1$, are also shown.}
    \label{fig:nc2_eos}
\end{figure}

The separation of medium-dependent contributions in the gap equations, together with a consistent treatment of divergent vacuum terms, has a direct impact on the resulting equation of state, as illustrated in Figure~\ref{fig:nc2_eos}. In order to quantify these effects, we evaluate a set of thermodynamic observables and analyze how different regularization prescriptions constrain and modify the thermodynamics of the system. The quantities considered include the pressure, the quark number density, the energy density, and the squared speed of sound, defined, respectively, as
\begin{align} 
P &= - [\Omega(\mu) - \Omega(\mu = 0)], \label{press}\\ 
n &= \frac{\partial P}{\partial \mu},\label{ndens}\\ 
\varepsilon &= - P + \mu n \label{EnDens}, \\ 
c^2_s &= \frac{\partial P}{\partial \varepsilon} \label{cs2}. 
\end{align}
{For} the traditionally regularized NJL model, employing either the three-dimensional-cutoff or Pauli–Villars schemes, the equation of state exhibits a similar qualitative behavior in both cases, with no substantial change in slope as the energy density $\varepsilon$ increases. The two curves remain close to each other, displaying only small deviations at larger values of $\varepsilon$.

A qualitatively different behavior emerges once the medium separation scheme is implemented. As shown in Figure~\ref{fig:nc2_eos}, the MSS prescription leads to a progressive softening of the equation of state with increasing energy density, highlighting the role of correctly isolating medium contributions. This demonstrates that the proper separation of medium-dependent terms is essential for obtaining a softer equation of state at moderate-to-high energy densities, significantly modifying the thermodynamic properties of quark matter even at asymptotically large values of the quark chemical potential, where convergence toward the conformal limit is expected.

\begin{figure}[b]
    \includegraphics[width=0.65\linewidth]{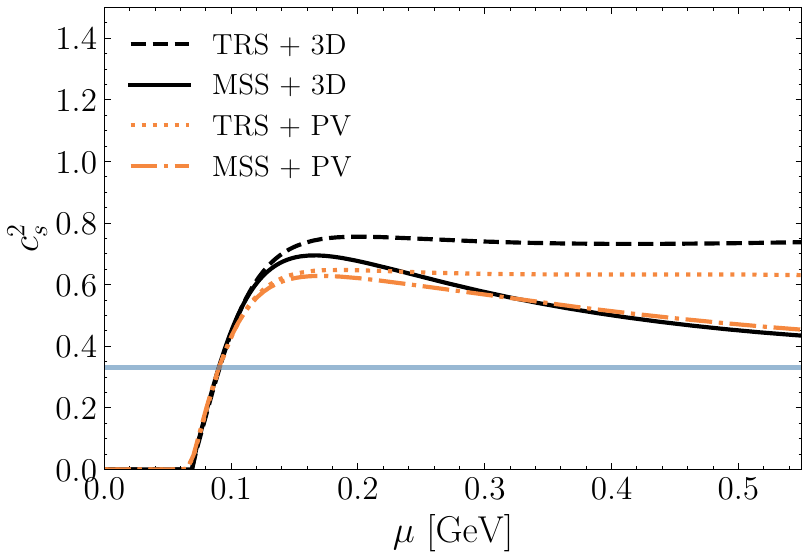}
    \caption{Speed of sound squared $c_s^2$ as a function of quark chemical potential $\mu$. The line styles follow the same convention as in Figure~\ref{fig:iso_mass}, and the light-blue solid line represents the conformal bound $c_s^2 = 1/3$.}
    \label{fig:nc2_cs2}
\end{figure}

This behavior becomes particularly evident when examining the squared speed of sound, $c_s^2$, as a function of $\mu$, as may be seen in Figure~\ref{fig:nc2_cs2} for all regularization schemes considered in this work. All the curves become nonzero at the onset of diquark condensation, $\mu = m_{\pi}/2$, and initially follow closely similar trajectories, violating the conformal bound at $\mu\approx 0.09$ GeV. The curves obtained using Pauli–Villars regularization begin to deviate from their three-dimensional-cutoff counterparts at $\mu\approx 0.12$ GeV. At slightly higher chemical potentials, $\mu\approx 0.14$ GeV, the TRS and MSS results within the Pauli--Villars scheme themselves start to separate.
Within Pauli–Villars regularization, the TRS prediction for the speed of sound reaches a plateau at an approximately constant value of $c_s^2\approx 0.61$. In contrast, the corresponding MSS result decreases steadily after attaining a similar maximum, eventually converging toward the conformal limit, as required for a physically consistent extrapolation at large values of $\mu$. The MSS result obtained with the three-dimensional cutoff displays a qualitatively similar peak structure, occurring at approximately the same chemical potential as in the Pauli–Villars case. {The peak behavior of $c_s^2$ was predicted by first-principle lattice QCD calculations~\cite{Itou:2025vcy}, where the squared speed of sound was calculated as a function of the quark chemical potential for $T=80$ MeV and $T= 40$ MeV. Our results displayed in Figure~\ref{fig:nc2_cs2} stand in strong agreement with LQCD predictions.} By contrast, the TRS curves, regardless of whether a three-dimensional-cutoff or Pauli–Villars regularization is employed, exhibit neither a pronounced peak nor any indication of convergence toward the conformal limit in the large $\mu$ limit.
{Finally, and most importantly, this distinctive thermodynamic behavior emerges within the conventional validity range of standard NJL model regularizations. As shown in Figure~\ref{fig:nc2_cs2}, the divergence in the speed of sound becomes apparent at chemical potential values well below both the $\Lambda_{\text{3D}}$ and $\Lambda_{\text{PV}}$ cutoffs. This confirms that the TRS significantly influences thermodynamic predictions, even within the chemical potential regimes where the model is typically considered reliable.}

As demonstrated throughout this section, medium contributions—often improperly regularized—play a decisive role in shaping the thermodynamic properties of dense quark matter in two-color QCD. Such improper treatment, while consistently violating the conformal bound, prevents the thermodynamic observables from converging toward their expected conformal behavior at asymptotically large chemical potentials. This shortcoming is most clearly manifested in the behavior of the speed of sound, which is expected to approach $c_s^2\to 1/3$ in the limit of high chemical potential.

A consistent regularization of the theory requires that only vacuum contributions be subjected to regularization, a condition that is naturally fulfilled within the medium separation scheme. By explicitly isolating medium-dependent terms, the MSS ensures that all the medium effects are properly incorporated, restoring the correct asymptotic behavior of thermodynamic quantities in the high-density regime of two-color QCD.

\section{Two-Flavor Color Superconductivity}
\label{sec:2sc+beta}

In this section, we briefly review the main aspects of the more realistic SU(2)$_f$ NJL model with scalar--pseudoscalar and scalar diquark channels. The corresponding Lagrangian density is given by~\cite{Klevansky:1992qe,Buballa:2003qv}:
\begin{align}
    \mathcal{L} &= \Bar{\psi}(i\gamma^{\mu} \partial_\mu - m_c + \mu \gamma^0)\psi + G_S[(\Bar{\psi}\psi)^2 + (\Bar{\psi}i\gamma^5\Vec{\tau}\psi)^2]\nonumber\\ 
    & + G_D [(\Bar{\psi}i\gamma^5\tau_2\lambda_A\psi_c)(\Bar{\psi}_ci\gamma^5\tau_2\lambda_A\psi)],
\end{align}
where $\psi_c = C \Bar{\psi}^T$ and $\Bar{\psi}_c = \psi^T C$ are charge-conjugate spinors, with $C = i \gamma^2 \gamma^0$ being the charge conjugation matrix, and $\vec{\tau} = (\tau_1,\tau_2,\tau_3)$ are the Pauli matrices in flavor space. The quark field $\psi \equiv \psi_{i,\alpha}$ with $i = 1,2$ and $\alpha = 1,2,3$ is a four-component spinor transforming as a flavor doublet and color triplet. The matrices $\lambda_A$ act in the color space and are antisymmetric. The current quark mass $m_c$ is taken to be diagonal, and the quark chemical potential is introduced through the diagonal $6\times6$ matrix $\mu=\left(\mu_{ur},\mu_{ug},\mu_{ub},\mu_{dr},\mu_{dg},\mu_{db}\right)$. In the isospin-symmetric limit, we assume $m_u = m_d\equiv m_c$ and $\mu_i \equiv \mu$. The parameters $G_S$ and $G_D$ denote the effective coupling constants in the scalar and diquark channels, respectively.
In accordance with the Pauli exclusion principle, the diquark condensate must be antisymmetric in Dirac, flavor, and color spaces. This requirement allows the condensate to be written in the form
$
\Delta \propto \varepsilon_{ij}\,\varepsilon^{\alpha\beta 3},
$
where the Latin indices $i,j$ label flavor degrees of freedom, while the Greek indices $\alpha,\beta$ denote color components. The fixed color index $3$~specifies the chosen orientation in color space. By convention, we choose the blue direction as the reference, implying that only the red and green quarks participate in the pairing, while the blue quark remains unpaired. Consequently, the relevant antisymmetric color operator is identified with $\lambda_A = \lambda_2$.  

Within the mean-field approximation, the Lagrangian density can then be rewritten in the form 
\begin{align}
    \mathcal{L} &= \Bar{\psi}(i\gamma^{\mu} \partial_{\mu} - M + \mu \gamma^0)\psi - \frac{1}{2} \Delta^*_{2\text{SC}}(i\Bar{\psi}_c \gamma^5 \tau_2 \lambda_2\psi) - \frac{1}{2} \Delta_{2\text{SC}}(i \Bar{\psi}\gamma^5 \tau_2 \lambda_2 \psi_c) \nonumber\\ 
    &- \frac{(M - m_0)^2}{4 G_S} - \frac{\Delta^*_{2\text{SC}} \Delta _{2\text{SC}}}{4 G_D}.
\end{align} 
{Here}, we define the effective quark mass $M$ and diquark condensates $\Delta_{\text{2SC}},\Delta_{\text{2SC}}^*$ as
\begin{align}
    &M = m_c  - 2G_S \sigma, \quad\quad\text{with}\quad \quad \sigma = \langle \Bar{\psi}\psi\rangle, \\
    & \Delta_{2\text{SC}} = - 2 G_D \langle \Bar{\psi}_c i \gamma^5\tau_2\lambda_2 \psi \rangle, \\ 
    & \Delta^*_{2\text{SC}} = - 2 G_D \langle \Bar{\psi}i\gamma^5 \tau_2 \lambda_2 \psi_c\rangle.
\end{align} 

In order for quark matter to exist in the core of compact objects, such as neutron stars, it is necessary to impose both $\beta$-equilibrium and charge neutrality. Under these constraints, the quark chemical potentials can be expressed as~\cite{Alford:2002kj,Steiner:2002gx,Huang:2002zd,Duarte:2018kfd,Yang:2025ach}: 
\begin{align}
    \mu_i = \mu - Q_e \mu_e + T_3 \mu_{3} + T_8 \mu_8,
\end{align} 
where $Q_e$, $T_3$, and $T_8$ are the generators of $U(1)_Q$, $U(1)_3$, and $U(1)_8$, respectively. Since diquark pairing involves only two colors, while the remaining one is unpaired and the remaining colors are degenerated, it is justified to set  $\mu_3 = 0$~\cite{Huang:2002zd, Ebert:2006bq,Yang:2025ach}. The quark chemical potentials, resolved by color and flavor, then read 
\begin{align}
    &\mu_{ur} = \mu_{ug} = \mu - \frac{2}{3}\mu_e + \frac{1}{3}\mu_8,\\ 
    &\mu_{dr} = \mu_{dg} = \mu + \frac{1}{3}\mu_e + \frac{1}{3}\mu_8,\\ 
    & \mu_{ub} = \mu - \frac{2}{3}\mu_e - \frac{2}{3}\mu_8,\\ 
    & \mu_{db} = \mu + \frac{1}{3} \mu_e - \frac{2}{3}\mu_8.
\end{align} 
{For} quarks of the same flavor, the color chemical potential $\mu_8$ accounts for the difference between the red and green sector and the blue sector, while $\mu_e$ controls the splitting between $u$ and $d$ quarks of the same color. Accordingly, we define the mean chemical potential, $\bar{\mu}$, for the paired quarks as  
\begin{align}
    \bar{\mu} = \frac{\mu_{ur} + \mu_{dg}}{2} = \frac{\mu_{ug} + \mu_{dr}}{2} = \mu - \frac{1}{6}\mu_e + \frac{1}{3}\mu_8,
\end{align} 
and the corresponding mismatch, $\delta \mu$, as 
\begin{align}
    \delta \mu = \frac{\mu_{dg} - \mu_{ur}}{2} = \frac{\mu_{dr} - \mu_{ug}}{2} = \frac{\mu_e}{2}.
\end{align}

The thermodynamical potential for electrically and color-neutral 2SC quark matter is given by 
\begin{align}
    \Omega &= \Omega_e + \Omega_0 + \frac{(M - m_c)^2}{4 G_S} + \frac{\Delta^2_{2\text{SC}}}{4 G_D} + \Omega_{\text{reg}} + \Omega_{\text{med}},
           \label{Omega2sc}
\end{align} 
where 
\begin{align}
    \Omega_{\text{reg}} =  - 2 \int_{0}^{\Lambda} \frac{dp \ \vec{p}^2}{2\pi^2} \left ( 2 E_p + 2 E^{+}_{\Delta} + 2E^{-}_{\Delta} \right ),
\end{align} 
\begin{align}
    \Omega_{\text{med}} &= -2\int_{0}^{\infty} \frac{dp  \ \vec{p}^2}{2\pi^2} \bigg\{ T \ln[1 + \exp(-\beta E^+_{ub})] + T \ln[1 + \exp(-\beta E^-_{ub})]\nonumber\\ 
           &+ T \ln[1 + \exp(-\beta E^+_{db})] + T \ln[1 + \exp(-\beta E^-_{db})]\nonumber\\
           &+ 2 T\ln[1+ \exp(-\beta E^{+}_{\Delta^+})]+ 2 T\ln[1+ \exp(-\beta E^{+}_{\Delta^-})] \nonumber\\
           & +  2 T\ln[1+ \exp(-\beta E^{-}_{\Delta^+})] +  2 T\ln[1+ \exp(-\beta E^{-}_{\Delta^-})] \bigg\}, 
\end{align} 
correspond to the regularized and medium contributions to the thermodynamic potential, respectively.

The corresponding dispersion relations are
\begin{align}
&E^{\pm}_{\Delta^{\pm}} = E^{\pm}_{\Delta} \pm \delta\mu;\quad  E_{\Delta}^{\pm} = \sqrt{(E_p \pm \bar{\mu})^2 + \Delta^2_{2\text{SC}}},  \\
&E^{\pm}_{ub} = E_p \pm \mu_{ub} ; \quad E_p = \sqrt{\vec{p}^2 + M^2}, \\ 
& E^{\pm}_{db} = E_p \pm \mu_{db}.
\end{align} 
{The} contribution $\Omega_e$ corresponds to a gas of massless free electrons and is given by: 
\begin{align}
    \Omega_e = -\frac{\mu_e^4}{12 \pi^2}.
\end{align} 
{The} vacuum term $\Omega_0$ in Equation~(\ref{Omega2sc}) is obtained by setting
 $\mu=\mu_e=\mu_8=\Delta = 0$ and fixing the effective quark mass to its vacuum value $M_0$, yielding 
\begin{align}
    \Omega_0 = -\frac{(M_0 - m_c)^2}{4 G_S} + 12 \int \frac{d^3 \vec{p}}{(2\pi)^3} \sqrt{\vec{p}^2 + M^2_0}.
\end{align}  
{The} thermodynamic potential contains all the information required to determine the equilibrium properties of quark matter. In the present framework, the physical thermodynamic potential that defines the pressure of quark matter, $\Omega = -P$, is obtained from Equation~(\ref{Omega2sc}) after determining the values of $\mu_8$, $\mu_e$, $M$, and $\Delta_{2\text{SC}}$ that satisfy the color and electric charge neutrality conditions,
\begin{align}
n_8 &\equiv -\frac{\partial \Omega}{\partial \mu_8} = 0, \\
n_Q &\equiv -\frac{\partial \Omega}{\partial \mu_e} = 0,
\end{align}
together with the gap equations,
\begin{align}
\frac{\partial \Omega}{\partial M} =\frac{\partial \Omega}{\partial \Delta_{2\text{SC}}} &= 0.
\end{align}  
{As} motivated earlier, we consider both the traditional regularization approach and the medium separation scheme. In this case, the explicit expressions for the thermodynamic potential and the gap equations are constructed following~\cite{Duarte:2018kfd}, based on the steps described in Section~\ref{sec:mss}.

The model parameters adopted for the three-color case are the same as those employed in Section~\ref{sec:isospin}, and summarized in Table~\ref{tab:iso_parameters}. In the diquark channel, the coupling constant $G_D$ is related to the scalar coupling $G_S$ through a Fierz transformation of the four-fermion interaction, yielding $G_D = 0.75G_S$~\cite{Buballa:2003qv}, which we adopt for the three-color case. Likewise, the thermodynamic quantities are defined following the same prescription introduced in Section~\ref{sec:qc2c}, see Equations (\ref{press})--(\ref{cs2}).

\begin{figure}[b]
    \includegraphics[width=0.65\linewidth]{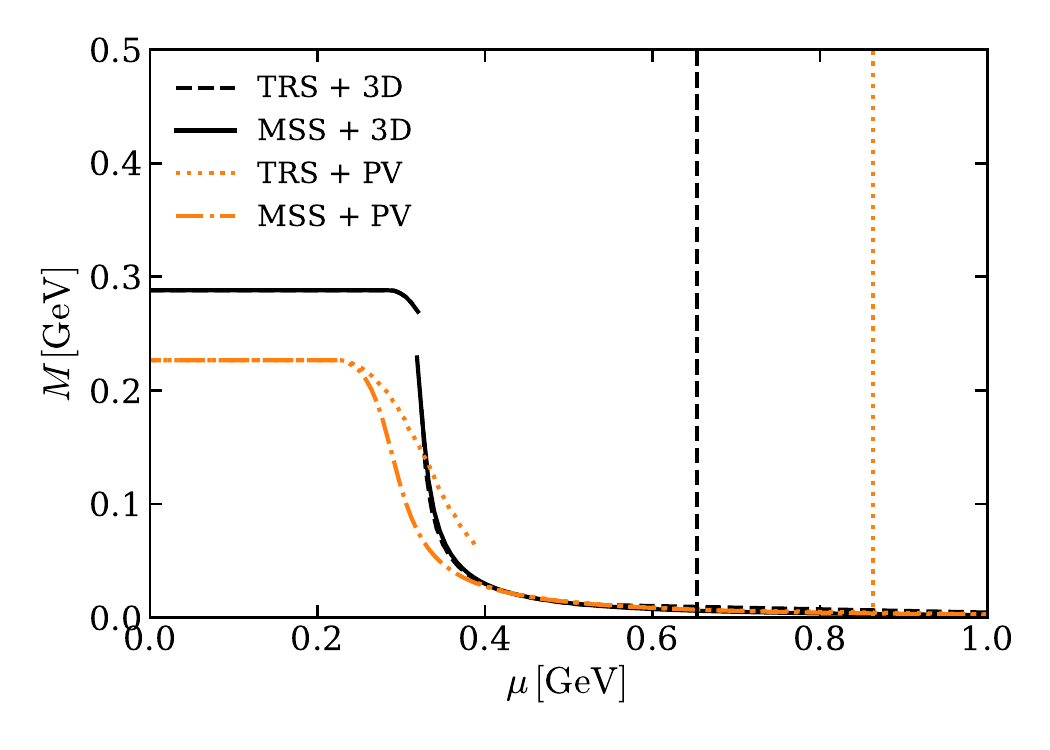}    
    \caption{{Effective} quark mass $M$ as a function of quark chemical potential $\mu$. {The vertical black dashed line indicates the scale $\Lambda_{\text{3D}}$, while the vertical orange dotted line represents the corresponding Pauli--Villars (PV) cutoff, $\Lambda_{\text{PV}}$.} The line styles follow the same convention as in Figure~\ref{fig:iso_mass}.}
    \label{figMass2SC}
\end{figure} 

Figure~\ref{figMass2SC} displays the effective quark mass $M$ as a function of the quark chemical potential $\mu$ for different regularization schemes and treatments of medium contributions. At low chemical potentials, $\mu \lesssim 0.30\,\mathrm{GeV}$, all curves exhibit a nearly constant plateau, whose value is determined by the vacuum parametrization: schemes employing a three-dimensional cutoff yield larger dynamical masses, while Pauli--Villars regularization leads to systematically smaller values. As the chemical potential increases, the effective mass decreases smoothly, signaling the gradual weakening of the scalar condensate and the onset of chiral symmetry restoration. Although the qualitative behavior is similar for all prescriptions, the precise chemical potential at which the mass drops more rapidly depends on the regularization scheme. At sufficiently high chemical potentials, $\mu \gtrsim 0.40\,\mathrm{GeV}$, all curves converge to small values governed mainly by the current quark mass, indicating that in this regime the effective mass becomes largely insensitive to both the vacuum regularization and the medium-separation procedure.

\begin{figure}[t]
    \includegraphics[width=0.65\linewidth]{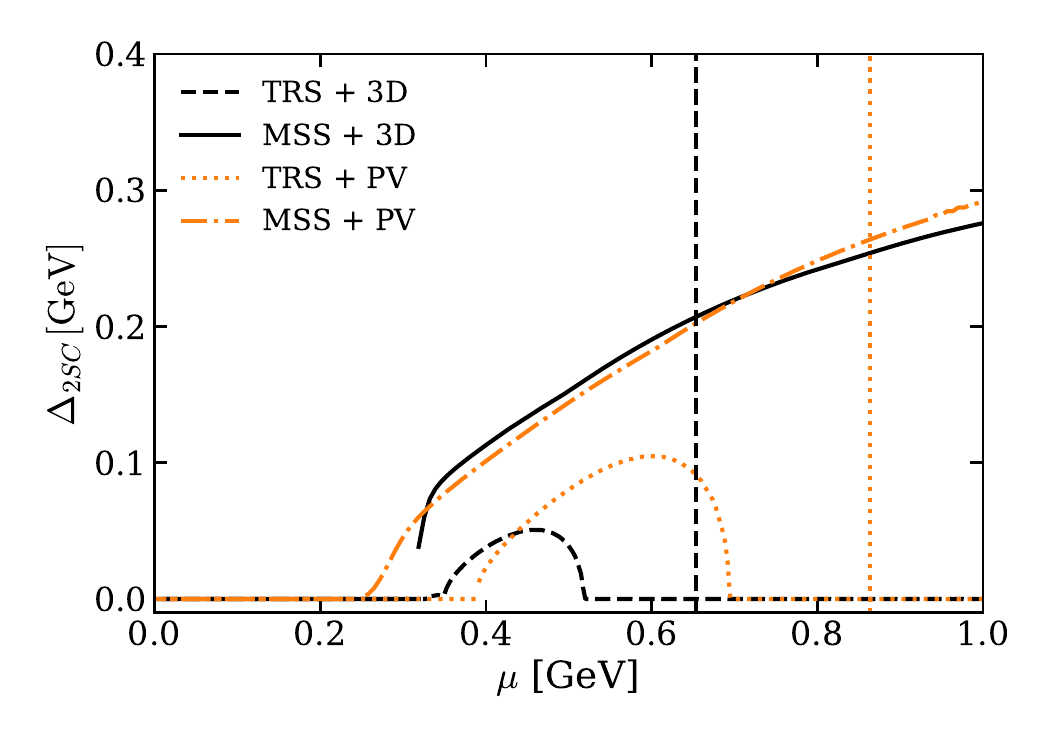}
    \caption{{Diquark} condensate $\Delta_{2\text{SC}}$ as a function of quark chemical potential $\mu$. {The vertical black dashed line indicates the scale $\Lambda_{\text{3D}}$, while the vertical orange dotted line represents the corresponding Pauli--Villars (PV) cutoff, $\Lambda_{\text{PV}}$.} The line styles follow the same convention as in Figure~\ref{fig:iso_mass}.}
    \label{figDelta2SC}
\end{figure}

\begin{figure}[b]
    \includegraphics[width=0.65\linewidth]{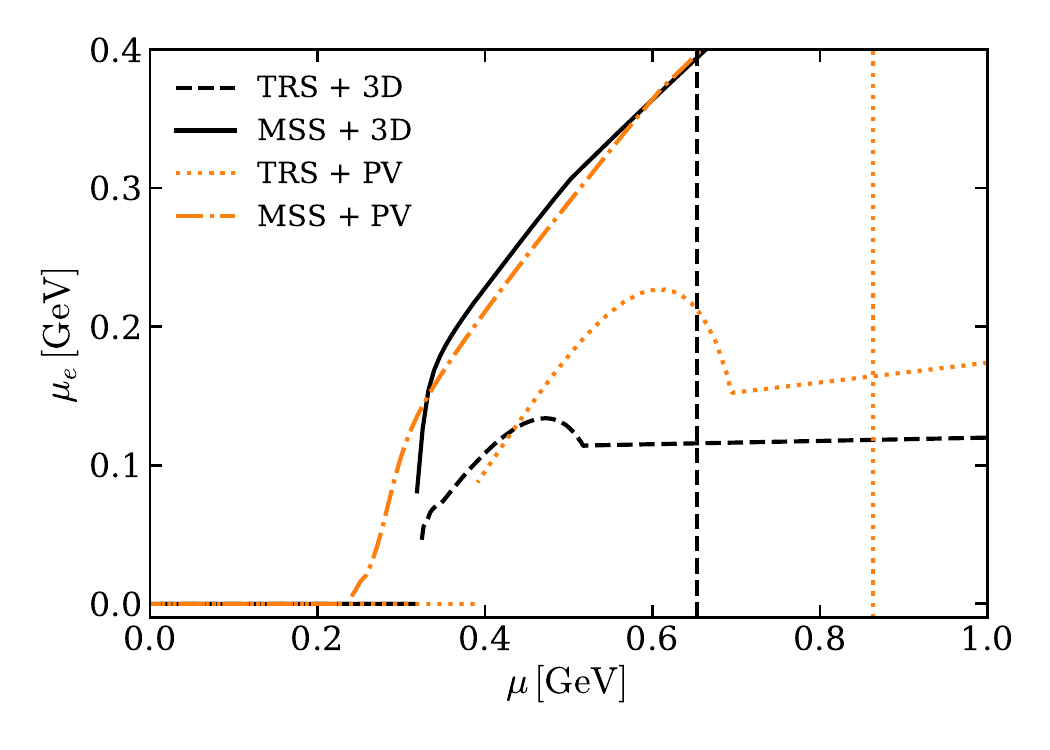}
    \caption{The electron chemical potential $\mu_e$ as a function of quark chemical potential $\mu$. {The vertical black dashed line indicates the scale $\Lambda_{\text{3D}}$, while the vertical orange dotted line represents the corresponding Pauli--Villars (PV) cutoff, $\Lambda_{\text{PV}}$.} The line styles follow the same convention as in Figure~\ref{fig:iso_mass}.}
    \label{figmue2SC}
\end{figure}

Figure~\ref{figDelta2SC} illustrates that, unlike the effective quark mass, the diquark condensate $\Delta_{2\text{SC}}$ is highly sensitive to the treatment of medium contributions. Within the TRS, both in the 3D-cutoff and Pauli--Villars implementations, the condensate exhibits a nonmonotonic behavior: after its onset, $\Delta_{2\text{SC}}$ initially grows but eventually decreases and vanishes at moderate chemical potentials. This feature is a well-known unphysical artifact of cutoff-based regularization schemes. In contrast, the MSS results are qualitatively different: $\Delta_{2\text{SC}}$ grows smoothly and monotonically with $\mu$, with no indication of decrease over the explored density range. This behavior reflects the proper separation between vacuum and medium contributions, which prevents medium-dependent integrals from being artificially suppressed. A comparison between 3D-cutoff and PV implementations within the MSS shows only mild quantitative differences, indicating that once medium divergences are treated consistently, the behavior of the 2SC gap becomes essentially regulator-independent and consistent with the expected physics of dense quark matter.

Figure~\ref{figmue2SC} displays the electron chemical potential $\mu_e$ as a function of the quark chemical potential $\mu$, illustrating how different regularization prescriptions affect the implementation of electric charge neutrality in 2SC quark matter. The results obtained within the MSS, using either the 3D-cutoff or the Pauli--Villars method, yield systematically larger values of $\mu_e$. This behavior arises because, in the MSS, medium-dependent contributions are integrated over the full momentum space, allowing the flavor imbalance between $u$ and $d$ quarks to develop in a natural way. In contrast, the TRS leads to a smoother, and in some cases, nonmonotonic behavior, most pronounced in the 3D-cutoff case. This feature can be traced back to the artificial mixing between the vacuum and medium terms induced by the uniform application of the cutoff. Overall, this figure shows that the MSS provides a more consistent and physically reliable description of electric neutrality in the 2SC phase, avoiding the artifacts inherent in the TRS.

\begin{figure}[b]
    \includegraphics[width=0.65\linewidth]{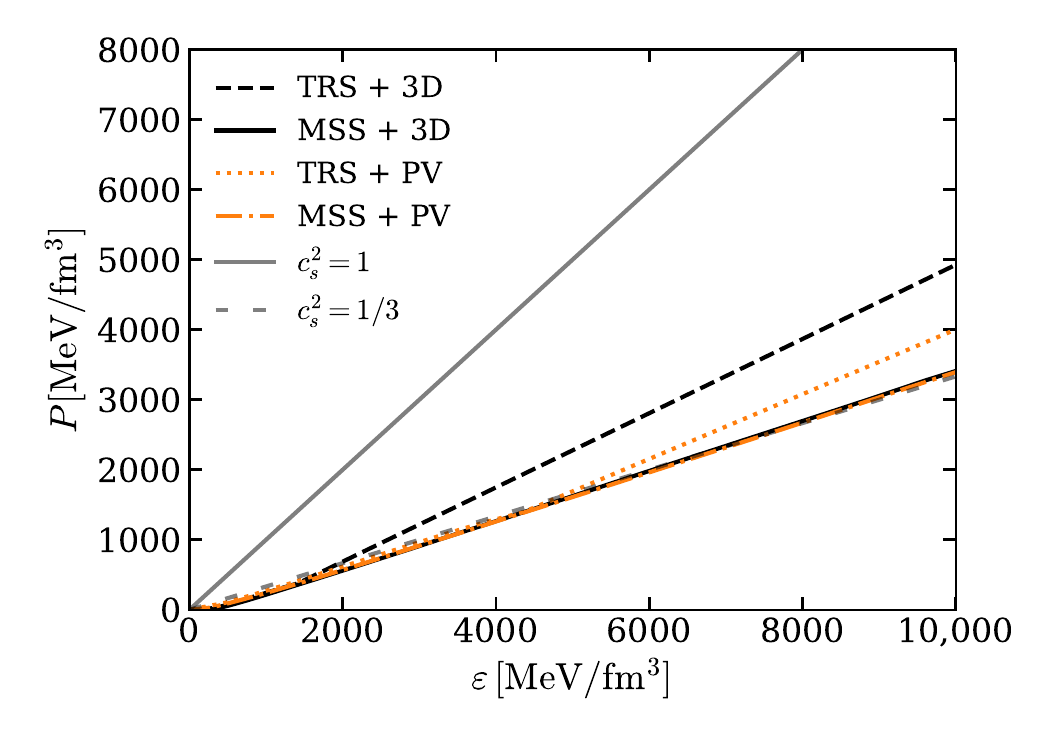}
    \caption{Equation 
 of state $P \times \varepsilon$ for 2-flavor QCD matter under stellar conditions. The line styles follow the same convention as in Figure~\ref{fig:iso_mass}, and guidelines corresponding to constant values of the speed of sound, $c_s^2 = 1/3$ and $c_s^2 = 1$, are also shown.}
    \label{figEOS2SC}
\end{figure}

The equation of state $P \times \varepsilon$ is shown in Figure~\ref{figEOS2SC}. The results show that the MSS yields a noticeably smoother equation of state at high chemical potentials when compared to the TRS at the same energy density. This behavior is related to the fact that, within the TRS employing a 3D cutoff, high-momentum contributions are artificially suppressed, resulting in an excessively stiff equation of state. When Pauli--Villars regularization is adopted, these cutoff artifacts are partially mitigated, and the TRS + PV curve becomes less stiff than its TRS + 3D counterpart, particularly at low and intermediate energy densities. Nevertheless, even in the PV case, the TRS remains stiffer than the MSS at high energy densities, reflecting the residual mixing between vacuum and medium contributions. In contrast, the MSS, whether implemented with a 3D-cutoff or Pauli--Villars regularization, consistently preserves medium effects. As a consequence, the difference between \mbox{MSS + 3D} and \mbox{MSS + PV} remains small, with both prescriptions yielding a softer and more physically reliable equation of state. Overall, the comparison between the MSS and TRS reinforces that only the MSS effectively avoids regulator-induced artifacts widely discussed in the~literature.

\begin{figure}[t]
    \includegraphics[width=0.65\linewidth]{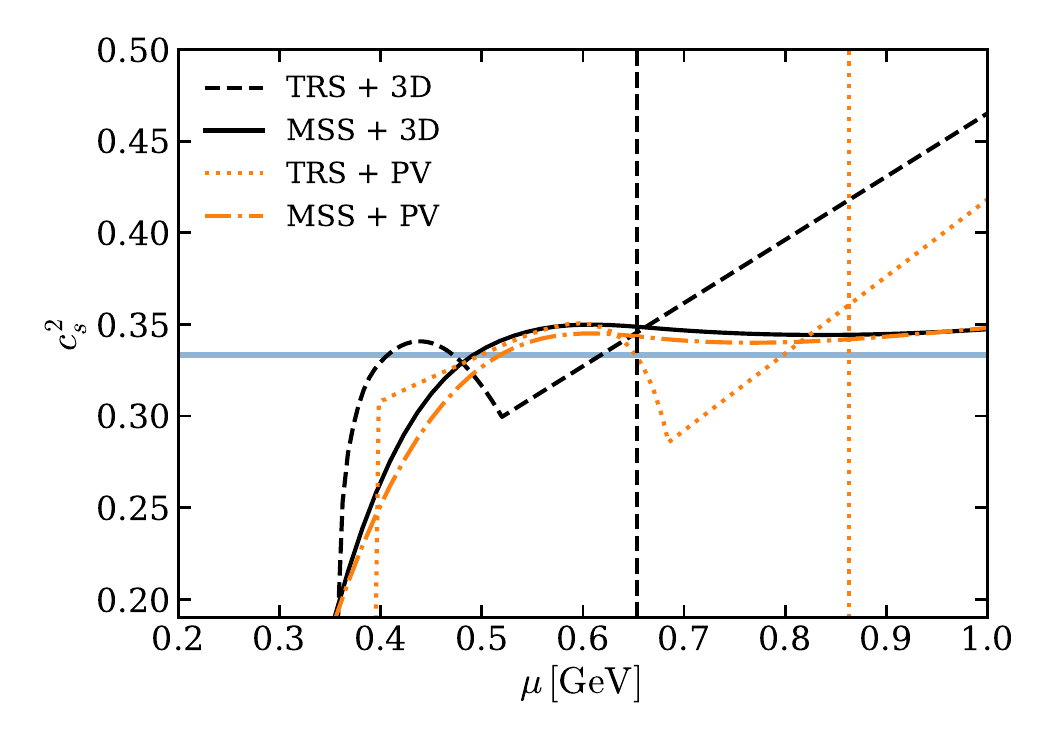}
    \caption{Speed of sound $c^2_s$ as a function of quark chemical potential $\mu$. The line styles follow the same convention as in Figure~\ref{fig:iso_mass}, and the light-blue solid line represents the conformal bound $c_s^2 = 1/3$. {The vertical black dashed line indicates the scale $\Lambda_{\text{3D}}$, while the vertical orange dotted line represents the corresponding Pauli--Villars (PV) cutoff, $\Lambda_{\text{PV}}$.} }
    \label{figcs22SC}
\end{figure} 

The squared speed of sound $c_s^2$ is shown in Figure~\ref{figcs22SC}. The results indicate that the TRS predictions, obtained using either the 3D-cutoff or the Pauli--Villars prescription, clearly exceed the conformal limit $c_s^2 = 1/3$, reflecting the artificial stiffening of the equation of state inherent to these schemes. The TRS + 3D case exhibits the most pronounced excess, as the uniform cutoff suppresses relevant medium contributions and renders the equation of state excessively stiff over a finite range of chemical potentials. Although the TRS + PV implementation mitigates part of these artifacts, it still produces a peak that rises above the conformal value due to the residual mixing between vacuum and medium terms. In contrast, the MSS results remain much more controlled: both MSS + 3D and MSS + PV display only moderate enhancements of $c_s^2$, with peaks that slightly exceed the conformal limit but are significantly reduced compared to the TRS cases. In particular, the MSS + PV curve is the smoothest among the four prescriptions, indicating a more consistent treatment of divergent integrals, while MSS + 3D shows a mild enhancement and approaches the conformal regime at higher chemical potentials. Overall, this comparison confirms that the MSS avoids the artificial stiffening characteristic of the TRS descriptions.

A central outcome of the present analysis is that the order of the chiral and diquark condensation transitions can be significantly affected by the presence of additional pairing channels. In two-color QCD in Section~\ref{sec:qc2c}, the onset transition corresponds to a Bose–Einstein condensation of gauge-invariant bosonic diquarks, for which the order parameter develops continuously and the low-energy dynamics is described by an effective bosonic theory, resulting in a genuine second-order phase transition. As previously mentioned, in that case, $G_D = G_S \equiv G$, and for physical parameter values, the transition remains of second order, even when the empirical quantities are rescaled by $N_c$. 
By contrast, in three-color QCD, diquark condensation in Section~\ref{sec:2sc+beta} arises from BCS-type pairing of fermionic quarks near the Fermi surface; since the associated condensate is not gauge invariant and couples to gluonic and density fluctuations, the transition is not governed by a simple critical fixed point and is therefore generically expected to be of first order or reduced to a crossover at phenomenologically relevant densities. In this case, the chiral and diquark condensation transitions are of first order when the Fierz transformation value of $G_D = 0.75 G_S$ is adopted. However, if the diquark coupling is treated as a free parameter, sufficiently large values of $G_D$ render the transitions of second order. A more detailed discussion on this topic can be found in Refs.~\cite{Sun:2007fc,Kitazawa:2007zs,Duarte:2017zdz}.

A meaningful comparison with perturbative QCD expectations can be made in light of the unified high-density analysis of Ref.~\cite{Fukushima:2024gmp}, where the 2SC phase is treated within the Cornwall--Jackiw--Tomboulis framework, including leading gap corrections. There, the squared speed of sound remains below the conformal value and approaches $c_s^2 = 1/3$ from below at asymptotically large $\mu$, without developing a peak above the conformal limit. In contrast, our MSS results display a different qualitative behavior: $c_s^2$ exhibits a pronounced enhancement above the conformal value in the physically relevant intermediate-density region before gradually relaxing toward $1/3$ at higher $\mu$. This behavior is consistent with recent nonperturbative approaches to 2SC matter, such as diquark--meson models~\cite{Andersen:2024qus} and approaches based on the Functional Renormalization Group (FRG)~\cite{Gholami:2025afm}, which also predict a significantly stiff equation of state in this regime. It is also in line with the findings of Ref.~\cite{Blaschke:2022egm}, where an effective description of color-superconducting quark matter matched to perturbative QCD asymptotics likewise yields a pronounced peak of $c_s^2$ above the conformal value at intermediate densities before approaching the asymptotic limit. Therefore, rather than reproducing the purely perturbative from-below trend, the MSS supports the picture of a strongly interacting 2SC phase with enhanced stiffness at intermediate densities while still remaining compatible with the expected convergence toward conformality at very high~density.

\section{Summary and Final Remarks} 
\label{Sec:remarks}

In this work, we examined the violation of the conformal limit for the speed of sound at finite density from the perspective of effective models, with particular emphasis on the role of regularization and on the consistent treatment of medium contributions. By applying the medium separation scheme to the NJL model, we analyzed three complementary settings: QCD at finite isospin density, two-color QCD, and two-flavor color superconductivity under stellar conditions, where either lattice QCD results or controlled theoretical expectations provided valuable benchmarks.

A central outcome of our analysis is that the qualitative behavior of key thermodynamic observables is largely governed by how medium effects are incorporated, rather than by the specific choice of vacuum regularization. Quantities dominated by vacuum physics, such as the effective quark mass, remain essentially insensitive to the MSS. In contrast, genuinely medium-driven observables, such as the pairing gaps, equations of state, and especially the speed of sound, are strongly affected by the improper regularization of medium contributions. Traditional schemes, which apply regulators uniformly to vacuum and medium terms, systematically suppress relevant physical effects, leading to unphysical artifacts such as the early disappearance of condensates, overly stiff equations of state, and persistent or uncontrolled violations of the conformal limit.

The peak structure observed in the speed of sound in all three scenarios analyzed arises from the proper vacuum regularization of the theory and from the way in which the regularization procedure affects the bosonic condensates and, consequently, the thermodynamics of strongly interacting matter. 
As shown in Figures~\ref{fig:iso_cond},~\ref{fig:nc2_Delta} and~\ref{figDelta2SC}, the value of each bosonic condensate is strongly influenced by the chosen regularization scheme. In particular, regularizing medium-dependent terms leads to a suppression of the condensate as the chemical potential approaches the natural cutoff scale of the theory. This suppression directly impacts the thermodynamic quantities, whose condensate contributions are diminished. 
The emergence of the peak structure in the speed of sound reflects the correct treatment of bosonic degrees of freedom:  near the Fermi surface, fermions that form bound bosonic pairs occupy opposite momentum states and therefore cease to contribute to the pressure as the chemical potential increases. In both two-color and asymmetric isospin cases, as the Fermi sphere expands, the system undergoes a BEC–BCS crossover, the fermionic interactions weaken, and the speed of sound gradually converges toward the conformal limit.

The MSS provides a natural resolution to these issues by enforcing the physically motivated requirement that only vacuum contributions must be regularized. Once this separation is implemented, condensates persist smoothly at high densities, equations of state soften in the moderate-to-high density regime, and the speed of sound develops a peak structure followed by convergence toward the conformal limit. This qualitative behavior is consistent with expectations from perturbative QCD at asymptotically large chemical potentials, where the conformal limit must be recovered, and with lattice QCD results and Gaussian-process analyses in regimes where nonperturbative effects remain important. In this sense, the MSS allows effective models to interpolate more reliably between low-density, lattice-accessible regions and the high-density domain where pQCD becomes applicable.

Finally, while effective models such as the NJL are not intended to provide quantitatively precise predictions at arbitrarily high densities, our results demonstrate that their qualitative reliability depends strongly on the consistent treatment of divergences. {In this sense, results at very large chemical potentials should be interpreted with care, since the NJL model gradually loses quantitative reliability as $\mu$ approaches (or exceeds) the characteristic ultraviolet scales of the regulator. Nevertheless, the appearance of a peak in $c_s^2$ takes place well within the regime where the model is reliable, whereas its subsequent qualitative behavior in approaching the conformal limit at $\mu > \Lambda$ should be interpreted as an extrapolation.} By eliminating regulator-induced artifacts and restoring the correct asymptotic behaviors determined by QCD, the MSS significantly extends the physical interpretability of these models. Properly regularized effective theories thus offer a meaningful and complementary framework to lattice QCD and perturbative approaches, helping to clarify the physical origin of conformal-limit violations and their resolution in dense QCD~matter.

\begin{acknowledgments}
This work was partially supported by Conselho Nacional de Desenvolvimento Cient\'{\i}fico e Tecnol\'ogico (CNPq), Grants No. 312032/2023-4, 402963/2024-5 and 445182/2024-5 (R.L.S.F.), No. 141270/2023-3 and 201300/2025-7 (B.S.L.); 
Fundação de Amparo à Pesquisa do Estado do Rio Grande do Sul (FAPERGS), Grants No. 24/2551-0001285-0 (R.L.S.F.), No. 23/2551-0000791-6, and No. 23/2551-0001591-9 (D.C.D);  
Coordena\c{c}\~ao de
Aperfei\c{c}oamento de Pessoal de N\'ivel Superior ---Brasil (CAPES) --- Finance Code 001 (F.X.A. and A.E.B.P.). The work is also part of the project
Instituto Nacional de Ci\^encia e Tecnologia --- F\'isica Nuclear e
Aplica\c{c}\~oes (INCT - FNA), Grant Nos. 464898/2014-5 and 408419/2024-5, and supported
by the Ser\-ra\-pi\-lhei\-ra Institute (grant number Serra --- 2211-42230).
\end{acknowledgments}

\end{document}